\documentclass[twocolumn,floatfix,times,trackchanges]{aastex63}

\def\sun{\hbox{$\odot$}}
\def\earth{\hbox{$\oplus$}}
\def\lesssim{\mathrel{\hbox{\rlap{\hbox{%
 \lower4pt\hbox{$\sim$}}}\hbox{$<$}}}}
\def\gtrsim{\mathrel{\hbox{\rlap{\hbox{%
 \lower4pt\hbox{$\sim$}}}\hbox{$>$}}}}

\def\arcmin{\hbox{$^\prime$}}
\def\arcs{\hbox{$^{\prime\prime}$}}

\def\farcs{\hbox{$.\!\!^{\prime\prime}$}}

\def\micron{\hbox{$\mu$m}}

\newcommand{\kmpers}{\mbox{~km~s$^{-1}$}}

\newcommand{\frest}{$f_\mathrm{rest}$}
\newcommand{\Aij}{$A_\mathrm{ij}$}
\newcommand{\Eu}{$E_\mathrm{u}$}
\newcommand{\gu}{$g_\mathrm{u}$}

\newcommand{\Msun}{$M_\odot$}

\usepackage{CJKutf8}
\usepackage{graphicx}
\usepackage{longtable} \LTcapwidth=\textwidth
\usepackage{multirow}
\usepackage{subfigure}
\usepackage{ulem}
\usepackage{soul}
\usepackage{lipsum, babel}
\usepackage{enumitem}
\usepackage{amsmath}
\usepackage{euscript}
\usepackage{verbatim}
\usepackage{tikz}

\accepted{\today}
\submitjournal{ApJ}

\shorttitle{Methanol in Orion Starless Cores}
\shortauthors{Hsu et al.}
\graphicspath{{./}{figures/}}

\begin{document}

\begin{CJK*}{UTF8}{bsmi}
\title{
Ubiquity of Methanol and its related Chemical Segregation in Orion Starless Cores: the ALMASOP Sample
}

\author[0000-0002-1369-1563]{Shih-Ying Hsu}
\email{seansyhsu@gmail.com}
\affiliation{Institute of Astronomy and Astrophysics, Academia Sinica, No.1, Sec. 4, Roosevelt Rd, Taipei 10617, Taiwan (R.O.C.)}

\author[0000-0012-3245-1234]{Sheng-Yuan Liu}
\email{syliu@asiaa.sinica.edu.tw}
\affiliation{Institute of Astronomy and Astrophysics, Academia Sinica, No.1, Sec. 4, Roosevelt Rd, Taipei 10617, Taiwan (R.O.C.)}

\author[0000-0001-8315-4248]{Xunchuan Liu}
\email{liuxunchuan001@gmail.com}
\affiliation{Shanghai Astronomical Observatory, Chinese Academy of Sciences, Shanghai 200030, PR China}

\author[0000-0001-8077-7095]{Pak Shing Li}
\affiliation{Shanghai Astronomical Observatory, Chinese Academy of Sciences, Shanghai 200030, PR China}

\author[0000-0002-8149-8546]{Ken'ichi Tatematsu}
\affiliation{Nobeyama Radio Observatory, National Astronomical Observatory of Japan, National Institutes of Natural Sciences, 462-2 Nobeyama, Minamimaki, Minamisaku, Nagano 384-1305, Japan}
\affiliation{Astronomical Science Program, The Graduate University for Advanced Studies, SOKENDAI,
2-21-1 Osawa, Mitaka, Tokyo 181-8588, Japan}

\author[0000-0001-9304-7884]{Naomi Hirano}
\affiliation{Institute of Astronomy and Astrophysics, Academia Sinica, No.1, Sec. 4, Roosevelt Rd, Taipei 10617, Taiwan (R.O.C.)}

\author[0000-0002-6868-4483]{Sheng-Jun Lin}
\affiliation{Institute of Astronomy and Astrophysics, Academia Sinica, No.1, Sec. 4, Roosevelt Rd, Taipei 10617, Taiwan (R.O.C.)}

\author[0000-0003-2011-8172]{Kee-Tae Kim}
\affiliation{Korea Astronomy and Space Science Institute (KASI), 776 Daedeokdae-ro, Yuseong-gu, Daejeon 34055, Republic of Korea}
\affiliation{University of Science and Technology, Korea (UST), 217 Gajeong-ro, Yuseong-gu, Daejeon 34113, Republic of Korea}

\author[0000-0003-1275-5251]{Shanghuo Li}
\affiliation{School of Astronomy and Space Science, Nanjing University, 163 Xianlin Avenue, Nanjing 210023, People’s Republic of China }
\affiliation{Key Laboratory of Modern Astronomy and Astrophysics (Nanjing University), Ministry of Education, Nanjing 210023, People’s Republic of China}

\author[0000-0002-5286-2564]{Tie Liu}
\affiliation{Shanghai Astronomical Observatory, Chinese Academy of Sciences, Shanghai 200030, PR China}

\author[0000-0002-4393-3463]{Dipen Sahu}
\affiliation{Physical Research laboratory, Navrangpura, Ahmedabad, Gujarat 380009, India}

\begin{abstract} 
Complex organic molecules (COMs) in starless cores provide critical insights into the early stages of star formation and prebiotic chemistry.
We present a chemical survey of 16 starless cores (including five prestellar cores) in the Orion A and B molecular clouds, targeting CH$_3$OH, N$_2$H$^+$, CCS, and c-C$_3$HD, using the Atacama Compact Array (ACA) and the Yebes 40-m telescope.
CH$_3$OH was detected toward all targets, confirming its ubiquity in starless cores, consistent with previous surveys in Taurus and Perseus.
ACA imaging shows that CH$_3$OH, CCS, and c-C$_3$HD generally trace the outer layers of the dense cores outlined by N$_2$H$^+$, each exhibiting distinct spatial distributions.
Meanwhile, Comparison with Yebes data reveals an extended, flattened CH$_3$OH component.
CCS and c-C$_3$HD tend to be detected or non-detected together across cores, while cores near dust-rich regions on a large scale often lack both, suggesting environmental influences linked to the interstellar radiation field.
Within individual cores, CCS typically resides in an outer layer relative to c-C$_3$HD.
Our findings underscore the importance of high-resolution studies for understanding the origins and spatial differentiation of COMs and carbon-chain molecules in cold, quiescent environments.
\end{abstract}

\keywords{astrochemistry --- ISM: molecules --- stars: formation and low-mass}

\section{Introduction}
\label{sec:Intro}

Complex organic molecules (COMs) in star-forming regions are of great interest due to their potential link to prebiotic chemistry in emerging planetary systems.
COMs, such as CH$_3$OH, are defined as saturated organic molecules composed of at least six atoms \citep{2009Herbst_COM_review}.
While COMs have been extensively studied in protostellar cores \citep[e.g.,][]{2020Belloche_CALYPSO,2020Hsu_ALMASOP,2020vanGelder_COMs,2021Yang_PEACHES,2022Hsu_ALMASOP,2022Bouvier_ORANGES}, 
their presence and behavior in starless cores remain less well studied.
Under the common nomenclature, a starless core is a dense condensation of gas and dust within a molecular cloud that lacks any protostellar object. 
Meanwhile, prestellar cores are considered a subset of starless cores that are gravitationally bound and expected to evolve into protostellar cores in the future \citep[e.g., ][]{2007DiFrancesco_review}.
In this paper, we use the term ``starless cores'' in a broader sense which includes those identified as prestellar cores.

The detection of COMs in starless cores has been reported for decades.
For example, \citet{1985Matthews_TMC1_L134N_COM} detected CH$_3$CHO in the starless cores TMC-1 and L134N.
\citet{1988Friberg_TMC1_L134N_B335_COM} later detected CH$_3$OH in the same regions.
\citet{2012Bacmann_L1689B_COM} reported the presence of CH$_3$CHO, CH$_3$OCH$_3$, and HCOOCH$_3$ in the starless core L1689B.
The other starless (including prestellar) cores reported to host detectable COMs include 
H-MM1 \citep{2020Harju_H-MM1_CH3OH}, 
IRAS 16293E \citep{2025Spezzano_IRAS16293E_COM,2025Scibelli_IRAS16293E_COM}, 
L429-C \citep{2023Taillard_L429-C_molec}, 
L1498 \citep{2006Tafalla_L1498_1517B}, 
L1544 \citep{2014Bizzocchi_L1544_CH3OH, 2014Vastel_L1544_COM, 2016Jimenez_L1544_COMs, 2016Spezzano_L1544_chem, 2017Spezzano_L1544_chem,2022Lin_L1544_CH3OH_HNCO}, 
L1517B \citep{2006Tafalla_L1498_1517B,2023Megias_L1517B_COM}, 
L1521E \citep{2019Nagy_L1521E_COM,2020Spezzano_survey_chem,2021Scibelli_L1521E_COM}, 
TMC-1 CP \citep{2015Soma_TMC1,2018Soma_TMC-1CP_COM}, 
TUKH122 \citep{2018Ohashi_TUKH122}. 
In recent years, COM surveys targeting a larger number of starless cores have begun to emerge.
For example, \citet{2020Scibelli_COM_Taurus} investigated COM compositions in 31 starless cores within the L1495–B218 filament of the Taurus molecular cloud. 
\citet{2024Scibelli_COM_Perseus} conducted a survey toward 35 starless cores in the Perseus cloud.
Interestingly, both of them found a 100\% detection rate of CH$_3$OH, suggesting the prevalence of COMs in starless cores.

In addition to investigating the occurrence of COMs, another important line of research focuses on their spatial distributions.
\citet{2006Tafalla_L1498_1517B} found that in L1498 and L1517B, CH$_3$OH exhibit a non-uniform, ring-like morphology that surrounds the dust continuum and N$_2$H$^+$ emission.
A similar morphology of CH$_3$OH has also been reported in, for example, 
L1544 \citep{2014Bizzocchi_L1544_CH3OH,2014Vastel_L1544_COM,2016Jimenez_L1544_COMs},
TUKH122 \citep{2018Ohashi_TUKH122}, 
H-MM1 \citep{2020Harju_H-MM1_CH3OH}, 
and five cores in the Orion cloud \citep{2025Hsu_ALMASOP_starless}. 
This ring-like structure is thought to be the result of molecular depletion in the densest regions of the cores \citep{2006Tafalla_L1498_1517B,2016Jimenez_L1544_COMs}. 
Subsequent studies showed that not only CH$_3$OH but also other COMs tend to be more abundant in the outer layers of the core, rather than at the continuum peak \citep{2021Jimenez_L1498_COMs, 2023Megias_L1517B_COM}. 
Finally, different distributions between CH$_3$OH and the carbon-chain molecules were also observed in L1544 \citep{2016Spezzano_L1544_chem, 2017Spezzano_L1544_chem} and toward a sample of six starless cores \citep[L1521E, HMM-1, OphD, B68, L429, and L694-2, ][]{2020Spezzano_survey_chem}.
Differences in exposure to the interstellar radiation field (ISRF) have been suggested to cause such spatial segregation \citep[e.g., ][]{2016Jimenez_L1544_COMs,2017Spezzano_L1544_chem,2020Spezzano_survey_chem}. 

The presence of gaseous COMs requires a desorption mechanism since COMs are believed to be formed in the icy mantles of dust grains \citep[e.g., ][]{2009Herbst_COM_review}. 
In protostellar cores, the commonly adopted mechanism is thermal desorption due to central protostar and shocks \citep[e.g., ][]{2002Velusamy_L1157_CH3OH,2008Arce_L1157-B1_COMs,2016Oya_IRAS16293-2422,2022Okoda_B335_chem,2023Hsu_ALMASOP,2024Hsu_HOPS87,2025Bouvier_HOPS409, 2025Hsu_G192}. 
However, in starless cores, the mechanism driving the presence of COM gas is still puzzling, despite the ample COM-detections in starless cores recently.
The plausible desorption mechanisms in this case include
reactive desorption \citep[][]{2007Garrod_reactive-desorption,2013Vasyunin_reactive-desorption,2017Vasyunin_reactive-desorption,2018Chuang_reactive-desorption,2020Jin_COM_reactive_desorption,2022Garrod_COM_reactive_desorption,2025Borshcheva_COM_reactive_desorption}, 
cosmic-ray-induced sputtering \citep{2020Dartois_CR-sputtering,2021Wakelam_CR_sputtering}, 
and grain-grain collision shocks \citep[][]{2001Dickens_TMC1, 2015Soma_TMC1, 2020Harju_H-MM1_CH3OH, 2022Lin_L1544_CH3OH_HNCO, 2022Kalvans_desorption, 2025Hsu_ALMASOP_starless}. 
Recently, \citet{2025Hsu_ALMASOP_starless} identified shock interfaces highlighted by the N$_2$H$^+$ and CH$_3$OH spatial and spectral correlation in several starless cores and proposed a ``turbulence-induced mass-assembly shock" scenario, bridging the ubiquitous turbulence activity and the common detection of CH$_3$OH in starless cores.

With the prevalence of COMs in starless cores readily revealed in the nearby low mass star forming Taurus and Perseus clouds, it is pertinent to expand such a census to the Orion molecular cloud.
While being at a farther distance of 400~pc (as compared to Taurus at 150~pc and Perseus cloud at 300~pc) \citep{2010Lombardi_2MASS_extinction_III,2011Lombardi_2MASS_extinction_IV,2018Zucker_Perseus_distance}, the Orion molecular clouds are known to experience stronger supersonic turbulence and more intense UV radiation fields \citep[e.g., ][]{2022Ha_SFR_turbulence,2022Xia_SFR_UV}.
Imaging observations similar to those in \citet{2025Hsu_ALMASOP_starless} will allow one to further investigate the critical role of COMs in elucidating the physical and chemical processes within the starless cores in a cloud environment that may be different from those in the Taurus and Perseus clouds.

\begin{deluxetable*}{lllllrlc}
\label{tab:coord}
\caption{
Information of the targets.
}
\tablehead{
\colhead{Name} & \colhead{Short Name} & \colhead{$\alpha$} & \colhead{$\delta$} & \colhead{Cloud} & \colhead{$v_\mathrm{LSR}$}  & \colhead{JCMT Name} & \colhead{Prestellar} \\
\colhead{} & \colhead{} & \colhead{(J2000)} & \colhead{(J2000)} & \colhead{} & \colhead{(km s$^{-1}$)}  & \colhead{} & \colhead{}
}
\startdata
G203.21-11.20E1 & G203.21E1 & 05:53:51.40 & $+$03:23:09.3 & Orion B & 10.62 & G203.21-11.20East1 & \\
G205.46-14.56M3 & G205.46M3 & 05:46:06.10 & $-$00:09:32.3 & Orion B & 10.16 & G205.46-14.56North1$^{\dagger}$ & \checkmark \\
G206.21-16.17N  & G206.21N  & 05:41:39.30 & $-$01:35:56.2 & Orion B & 9.90 & G206.21-16.17North & \\
G206.21-16.17S  & G206.21S  & 05:41:36.10 & $-$01:37:47.6 & Orion B & 9.50 & G206.21-16.17South & \checkmark \\
G209.29-19.65N1 & G209.29N1 & 05:35:00.80 & $-$05:39:57.7 & Orion A & 8.44 & G209.29-19.65North1 & \checkmark \\
G209.29-19.65S2 & G209.29S2 & 05:34:53.80 & $-$05:46:21.6 & Orion A & 7.44 & G209.29-19.65South2 & \\
G209.55-19.68N2 & G209.55N2 & 05:35:07.50 & $-$05:56:46.4 & Orion A & 8.15 & G209.55-19.68North2 & \\
G209.77-19.40E3 & G209.77E3 & 05:36:36.00 & $-$06:02:40.2 & Orion A & 7.76 & G209.77-19.40East3 & \\
G209.94-19.52N  & G209.94N  & 05:36:11.40 & $-$06:10:44.8 & Orion A & 8.17 & G209.94-19.52North & \checkmark \\
G210.37-19.53N  & G210.37N  & 05:36:55.20 & $-$06:34:35.2 & Orion A & 6.41 & G210.37-19.53North & \\
G210.82-19.47N2 & G210.82N2 & 05:38:00.00 & $-$06:57:17.5 & Orion A & 5.20 & G210.82-19.47North2 & \\
G211.16-19.33N4 & G211.16N4 & 05:38:56.10 & $-$07:11:27.9 & Orion A & 4.45 & G211.16-19.33North4 & \\
G211.16-19.33N5 & G211.16N5 & 05:38:46.00 & $-$07:10:41.9 & Orion A & 4.27 & G211.16-19.33North5 & \\
G212.10-19.15N1 & G212.10N1 & 05:41:21.30 & $-$07:52:26.9 & Orion A & 3.99 & G212.10-19.15North1 & \checkmark
\enddata
\tablecomments{
The coordinates are based on the continuum peak in our 3~mm observations. 
The only exception is G211.16-19.33N5 (G211.16N5) due to its nearby source HOPS-135 (05h38m45.347s, -07d10m56.04s). 
As a result, the coordinate of it here is the targeted coordinate, which was based on the ACA images of \citet{2020Dutta_ALMASOP}. 
The ``JCMT Name'' is the name of the source used in \citet{2018Yi_PGCC_SCUBA2_II}. 
In column ``Prestellar,'' the ``\checkmark'' marks sources identified as prestellar cores by \citet{2021Sahu_ALMASOP_presstellar}.
}
\end{deluxetable*}

In this study, we present emission maps of N$_2$H$^+$, CH$_3$OH, c-C$_3$HD, and CCS toward 16 starless cores in the Orion A and B molecular clouds, obtained with the Atacama Compact Array (ACA).
We also conducted observations with the Yebes 40-m telescope, allowing a comparison between single-dish and interferometric measurements.
Section~\ref{sec:Obs} describes the observational setups and data acquisition.
In Section~\ref{sec:results:ACA}, we characterize the general morphologies revealed in the ACA images.
Section~\ref{sec:results:Yebes} presents a comparison of CH$_3$OH column densities derived from the ACA and Yebes data.
In Section~\ref{sec:Disc} and \ref{sec:Conclusions}, we discuss the implications of our findings and summarizes our main results, respectively.
In addition, we present in Appendix~\ref{appx:Obs} additional detailed observational information and in Appendix~\ref{appx:overview} the overview of individual sources. 


\section{Methods}
\label{sec:Obs}

\subsection{Sample Selection}

In this study, we investigated 16 cold cores drawn from the ALMA Survey of Orion PGCCs (ALMASOP) project \citep{2020Dutta_ALMASOP}. 
The Planck Galactic Cold Clump (PGCC) catalogue provides an all-sky inventory of cold (10–20 K), dense clumps characterized by molecular hydrogen column densities of $N(H_2) > 10^{20}$ cm$^{-2}$ at an angular resolution of 5\arcmin \citep{2016Planck_PGCC}. 
Based on follow-up observations with the James Clerk Maxwell Telescope (JCMT) using its Submillimetre Common User Bolometer Array-2 (SCUBA-2) instrument, 119 dense cores were identified by their 850~\micron (dust continuum) emission within 96 PGCCs located in the Orion A, Orion B, and $\lambda$ Orionis clouds \citep{2018Yi_PGCC_SCUBA2_II}.
From these 119 cores, the ALMASOP project selected 72 compact and high-density sources and ultimately cataloged 23 starless cores and 56 protostellar cores \citep{2020Dutta_ALMASOP}.

Our sample consists of 16 starless cores chosen from this ALMASOP catalogue, selected for their compact distribution (within $\sim$2,000~au) in the 7-m ACA array data. 
All these cores were classified as starless by \citet{2018Yi_PGCC_SCUBA2_II} based on the absence of detections in the WISE bands (3.4–22~$\micron$).
In addition, all the cores exhibit strong N$_2$D$^+$ emission, indicative of their cold environments \citep{2020Kim_PGCC_45m}.
A summary of the basic properties of these 16 targets is provided in Table~\ref{tab:coord}.
In addition, an overview of the targets is in Appendix~\ref{appx:overview:lit}. 
These cores are distributed in the Orion A or B clouds with a distance of $\sim400$~pc \citep{2011Lombardi_2MASS_extinction_IV,2018Kounkel_APOGEE2}. 
The core masses range from $0.15$ to $3.52$ \Msun, reported by \citet{2018Yi_PGCC_SCUBA2_II} based on their 850~\micron\ dust continuum emission. 
Among the 16 cores, five of them show compact structures and extremely high peak densities (2-−8 $\times 10^7$ cm$^{-3}$) and are thus classified as prestellar cores \citep{2021Sahu_ALMASOP_presstellar}.
The remaining eleven exhibit flattened structures and lack continuum detections in the 12-m data. 

\begin{deluxetable}{lrrrrrl}
\setlength{\tabcolsep}{2pt}
\caption{\label{tab:molec_info}The molecular transitions in this report. }
\tablehead{\colhead{Formula} & \colhead{\frest} & \colhead{\Eu} & \colhead{\gu} & \colhead{\Aij} & \colhead{Quantum Numbers} \\
\colhead{} & \colhead{(GHz)} & \colhead{(K)} & \colhead{} & \colhead{(s$^{-1}$)} }
\startdata
\hline\multicolumn{6}{c}{ACA}\\\hline
N$_2$H$^+$ &  93.1734 & 4.47  & 27 & $3.6287E{-5}$ & $J=1-0$ $^\dagger$ \\ 
CCS        &  93.8701 & 19.89 & 17 & $3.7440E{-5}$ & $J_N=8_7-7_6$ \\ 
CH$_3$OH   &  96.7414 & 6.96  & 20 & $3.4075E{-6}$ & A, $2_{(0, 2)}-1_{(0, 1)}$ \\ 
c-C$_3$HD  & 107.4237 & 10.88 & 21 & $4.4652E{-5}$ & $3_{(1, 3)}-2_{(0, 2)}$ \\ 
\hline
\multicolumn{6}{c}{Yebes}\\
\hline
CH$_3$OH   &  48.3725 &  2.32 & 12 & $3.5502E{-7}$ & A, $1_{(0, 1)}-0_{(0, 0)}$ \\ 
\enddata
\tablecomments{\Eu\ is the upper energy. \frest\ is the rest frequency. \Eu\ is the upper state energy. \gu\ is the upper state degeneracy. 
$^\dagger$ The listed N$_2$H$^+$ transition contains the three hyperfine groups ($F_1=0-1$, $2-1$, $1-1$). We focus on the $F_1=0-1$ hyperfine group at 93.176~GHz and, under the assumption of optically thin emission, we multiplied it by a factor of nine when deriving the column densities.
}
\tablerefs{
The data were downloaded from the Splatalogue\footnote{\url{https://splatalogue.online/}} with their original references listed as follows. 
N$_2$H$^+$: \citet{2009Pagani_trans_N2H+}; 
CCS: \citet{1990Yamamoto_trans_CCS}; 
CH$_3$OH: \citet{1997Xu_trans_CH3OH}; 
c-C$_3$HD: \citet{1987Bogey_trans_C3H2}. 
}
\end{deluxetable}

\subsection{Observations}

\subsubsection{ACA Band~3 (3~mm) Observations}
\label{sec:Obs:ACA}

The majority of this study is based on data obtained with the Atacama Compact Array (ACA) during ALMA Cycle~8 under Project ID \#2021.2.00094.S (PI: Sheng-Yuan Liu), using Band~3 (3~mm) observations.
The observations were carried out between July 17, 2022, and April 17, 2023.
The executions were divided into two Scheduling Blocks (SBs): the first targeting 12 fields in the Orion A cloud, and the second targeting four fields in the Orion B cloud.
The unprojected baseline lengths ranged from 3 to 16 k$\lambda$.
The maximum recoverable scale of the observations was 60\arcsec, and the final images achieved an angular resolution of approximately 20\arcsec.

The receivers captured data across 12 spectral windows (SPWs.)
In this study, we focus on four transitions, N$_2$H$^+$ $J=1-0$, CCS $J = 8-7, N = 7-6$, CH$_3$OH $2_{0, 2}-1_{0, 1}$ ($E_\mathrm{u}=$6.96 K, hereafter CH$_3$OH-7K transition), and c-C$_3$HD $3_{1, 3}-2_{0, 2}$ transitions. 
The detailed transition parameters are listed in Table \ref{tab:molec_info}. 
For N$_2$H$^+$ $J=1-0$, we focus only on the hyperfine line group at 93.176 GHz ($F_1=0-1$) because (1) its hyperfine components are indistinguishable at our spectral resolution (61 kHz or 0.18 \kmpers), minimizing the effects of line blending in the channel maps, and (2) this group has the weakest intrinsic line strength of the three, making the emission comparatively less optically thick.
The corresponding SPWs for the four molecules are \#16, \#18, \#26, and \#42. 
Please refer to Table~\ref{tab:appx:spw} for the center frequencies, bandwidths, and the resolutions. 

\begin{figure}[htb!]
\centering
\includegraphics[width=.9\linewidth]{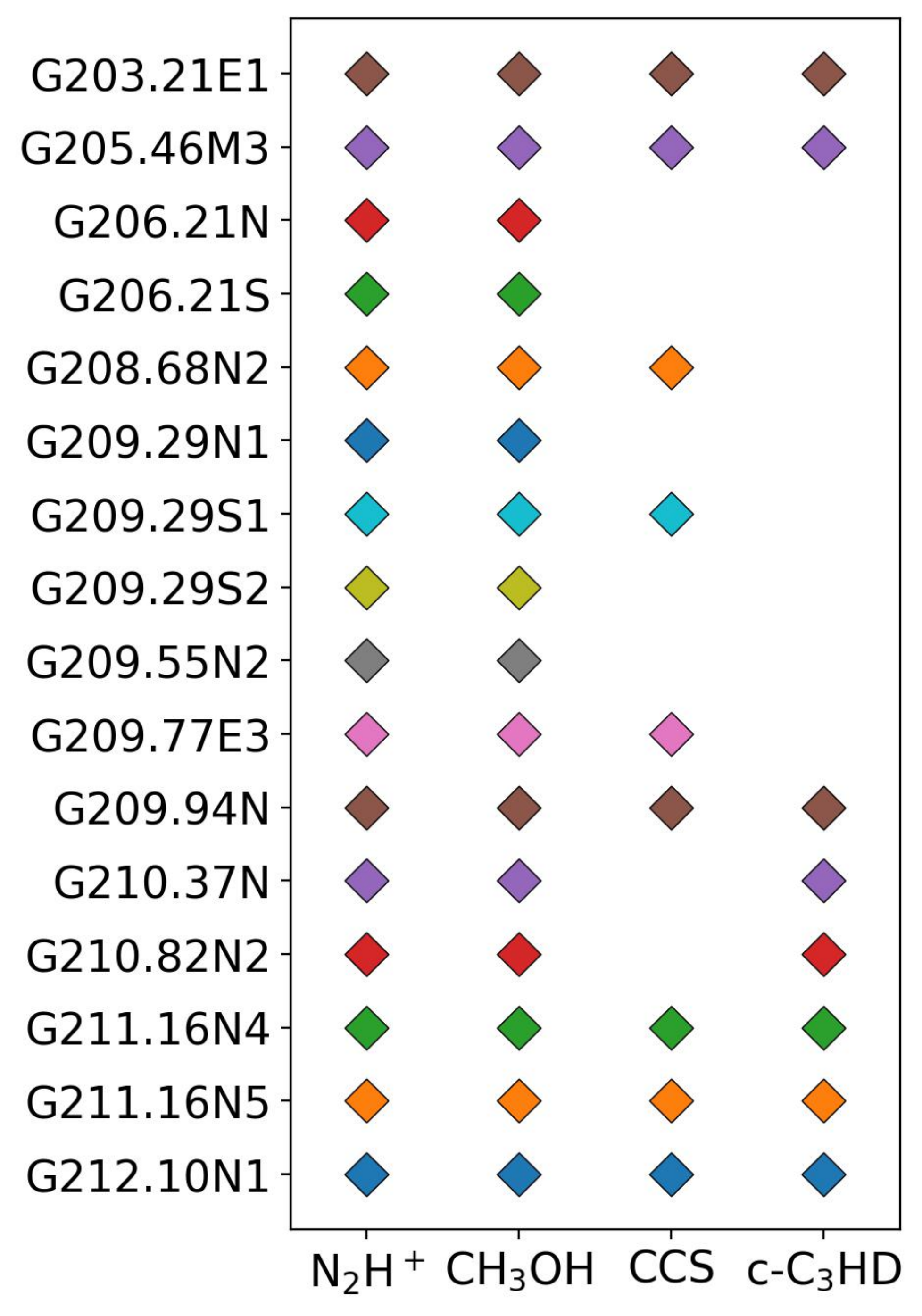}
\caption{\label{fig:scatter_detection} The ACA detection statistics of this study. 
}
\end{figure}

Imaging was performed using the \texttt{tclean} task in CASA 6.5.4 \citep{casa:2022}.
The Briggs weighting with a robust parameter of 2.0 was applied for better illustrating the extended emission and improving the sensitivities.
The imaging thresholds were 1.2~mJy for the full-band ($\sim$0.857~GHz) continuum and 100~mJy for individual channels (with a spectral resolution of 61~kHz per channel).
The final images achieved a (geometric mean) beam size of $\sim$14\arcs\ for the continuum and $\sim$18\arcs\ for all datacubes, respectively.
The resulting sensitivities were 0.5~mJy beam$^{-1}$ (0.2 mK) for the continuum and 25~mJy beam$^{-1}$ (10 mK) for each channel. 
Tables~\ref{tab:appx:obs} and \ref{tab:appx:noise} show the beam parameters and noise levels for each source and SPWs. 

\subsubsection{Yebes Observation at Q Band}
\label{sec:Obs:Yebes}

This study also utilized data obtained from the Yebes 40~m telescope under Project ID 22A010 (PI: Liu Xunchuan) in the Q band (31.00–50.50GHz).
This program targeted all 23 starless cores catalogued in the ALMASOP project, though the present study only uses those also observed with the ACA.
The spectral resolution was 38~kHz.
Each target was observed using the dual linear polarization Q-band receiver in frequency-switching mode with a standard throw of 10.52~MHz.
The integration time per source was 180 minutes.
Data reduction was performed using the \texttt{CLASS} program of the \texttt{GILDAS}\footnote{\url{https://www.iram.fr/IRAMFR/GILDAS/}} package \citep{2005GILDAS}. 

We focus on the CH$_3$OH $1_{(0,1)}$–$0_{(0,0)}$ transition at 48.372456 GHz (with an upper-level energy of 2.69~K; hereafter, the CH$_3$OH-3K transition) at which frequency the spectral resolution of 38~kHz corresponds to a velocity resolution of $\sim$0.24 \kmpers. 
The angular resolution is approximately 36\arcs\ at 48.4~GHz \citep{2021Tercero_Yebes}.
Table \ref{tab:appx:noise} shows the noise levels for each source. 

We conducted baseline subtraction for each source and extracted the spectrum centered at the expected line frequency with a 5~MHz window.
The typical noise level in the extracted spectra is $\sim$10~mK.


\section{Detection and Morphology revealed by Interferometry}
\label{sec:results:ACA}

\subsection{Detection}

We applied the following processes for identifying the molecular detections. 
First, we adopted the datacube without primary beam correction so that the voxels across the entire image have similar noise levels. 
A molecule is considered detected if any voxel has a flux density above five times the noise level within a sub-cube centered on the target position within a circular region of 25\arcsec\ (10,000 au) and centered on the source $v_\mathrm{LSR}$ covering $\pm$3 km s$^{-1}$ in velocity.

Figure~\ref{fig:scatter_detection} shows the detection statistics for each species toward each source.
As shown in Figure~\ref{fig:scatter_detection}, we detected N$_2$H$^+$ and CH$_3$OH toward all the sources. 
On the other hand, we detected CCS and c-C$_3$HD toward only 10 and 8 among 16 sources, respectively. 
Seven of these sources have both CCS and c-C$_3$HD detections together (G203.21E1, G205.46M3, G209.94N, G210.82N2, G211.16N4, G211.16N5, and G212.10N1).

\begin{figure}[htb!]
\centering
\includegraphics[width=.99\linewidth]{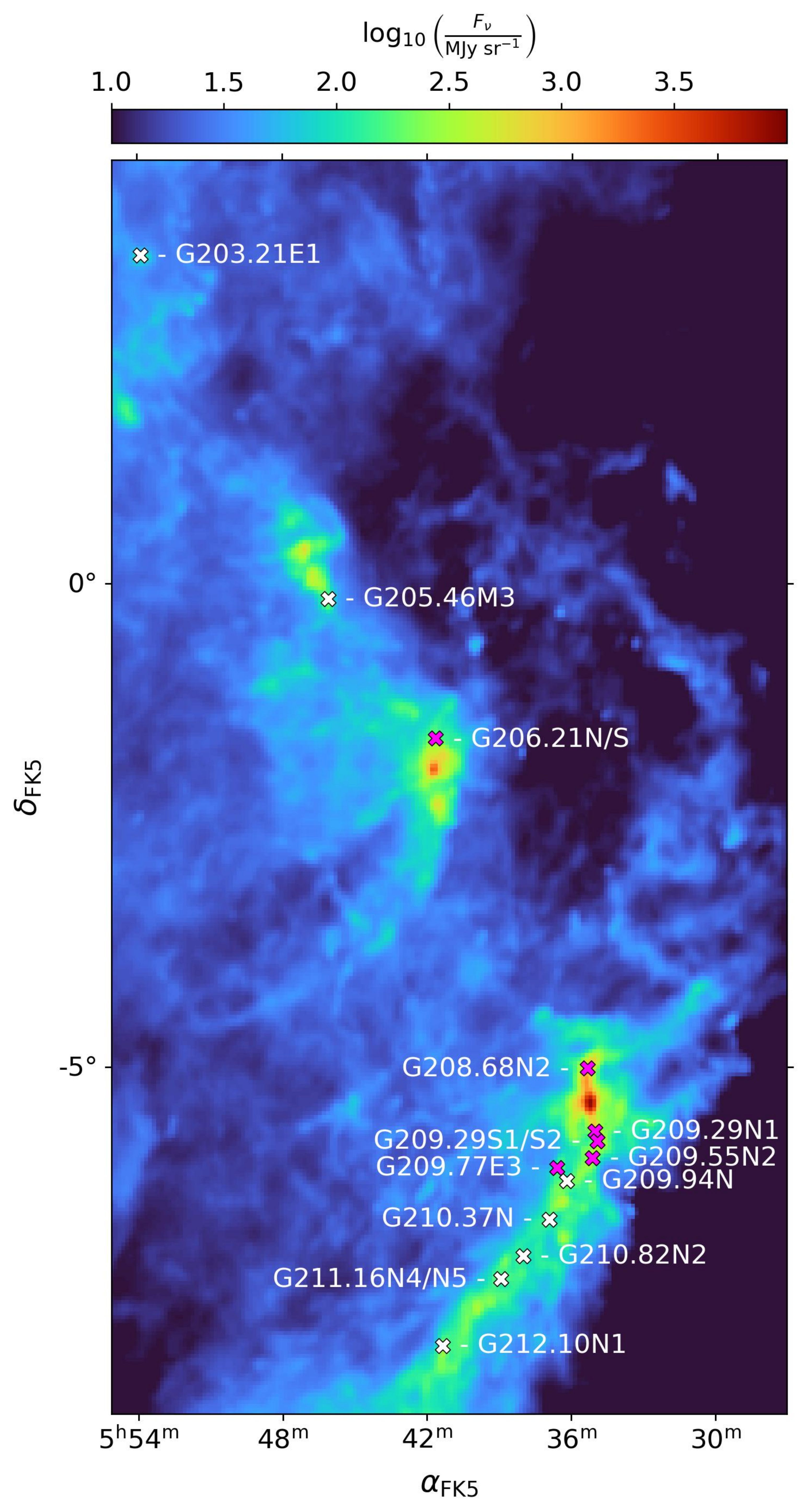}
\caption{\label{fig:scatter_OMC_857GHz} The Planck 857~GHz (350~\micron) map of the Orion A and B clouds covering our targets. 
Each marker represents one clump in this study. 
The white and magenta colors illustrate the detection and non-detection of c-C$_3$HD, respectively. 
The white texts label the names of the clumps as well as the cores within. 
}
\end{figure}

\begin{figure*}[htb!]
\centering
\includegraphics[width=.9\linewidth]{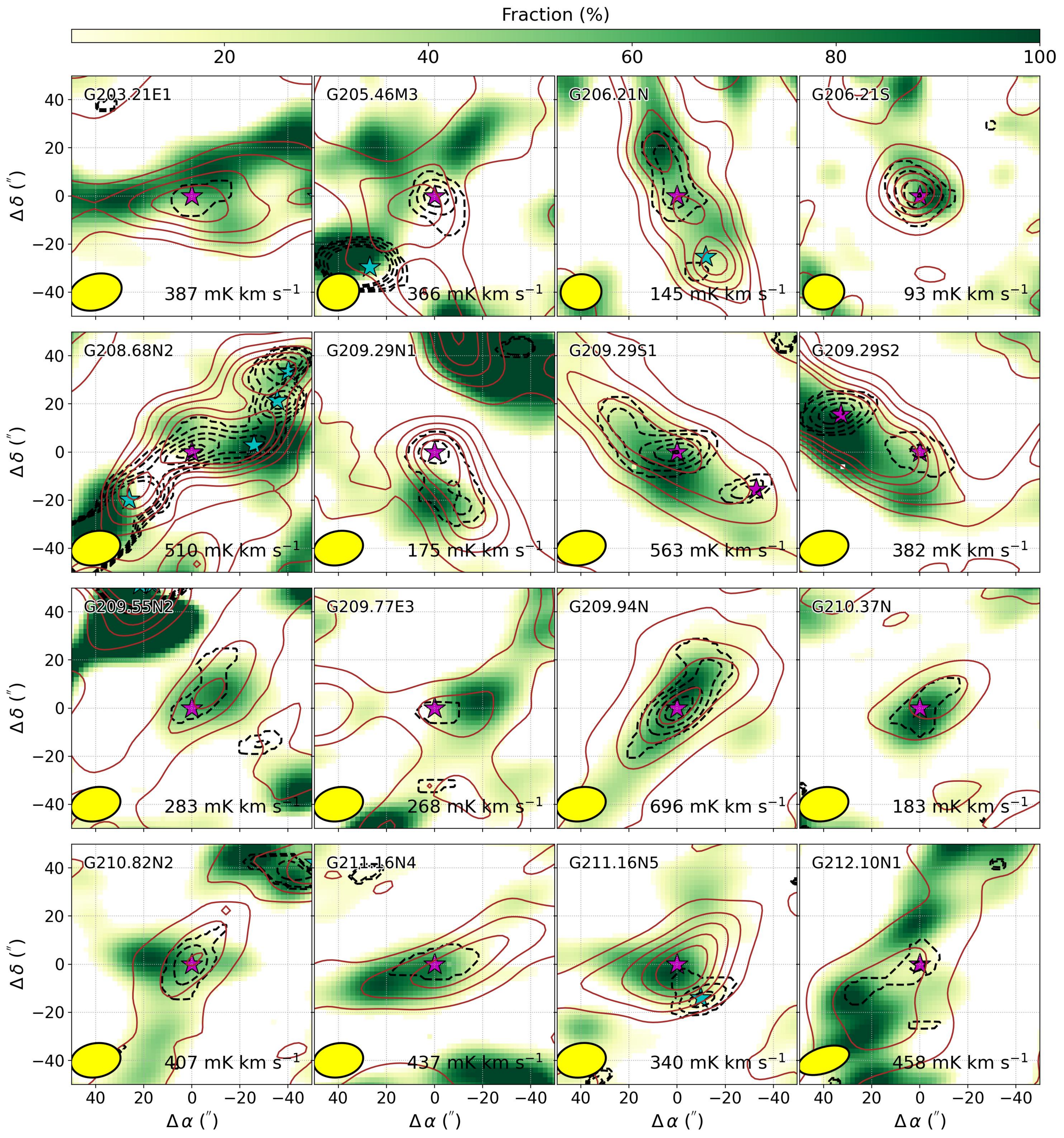}
\caption{\label{fig:mom0_main} The integrated intensity images of CH$_3$OH (rasters) overlaid with that of N$_2$H$^+$ (brown solid contours) and 3~mm dust continuum (black dashed contours). 
The velocity range for deriving the integrated intensity images was fixed to $v_\mathrm{LSR}\pm1$ km s$^{-1}$. 
The yellow ellipse at the lower left corner of each panel shows the synthesized beam of the rasters achieved by the ACA.
The text at the lower right corner of each panel corresponds to the peak value for the global color bar.
For five extremely dense starless cores, G205.46M3, G208.68N2, G209.94N, G209.29S1, and G212.10N1, the N$_2$H$^+$ contour levels are set at [5, 55, 105, 155, 205]$\sigma$; for the remaining 11 cores, the contour levels are [5, 25, 45, 65, 85]$\sigma$ with $\sigma\sim7.5$ mK km s$^{-1}$.
The dust continuum contours are plotted at levels of [3, 6, 9, 12, 15] of the continuum noise levels, which are shown in Table~\ref{tab:appx:noise}. 
The magenta markers indicate the sources included in this study, with the target coordinate as the reference; the corresponding coordinates are listed in Table~\ref{tab:coord}.
The cyan markers indicate nearby sources reported by literature, which can be found in the overview description of each source in Appendix~\ref{appx:overview:mom0}. 
}
\end{figure*}

We found tentative evidence for environmental effects influencing the chemical composition on large scales.
Several targets in our sample belong to the same PGCC, hence likely sharing similar environments. 
The cores within the same PGCCs have the same number in their short name, including G206.21N/S, G209.29S1/S2, and G211.16N4/N5.
As shown in Figure~\ref{fig:scatter_detection}, starless cores within the same PGCC tend to show generally consistent detections or non-detections of carbon-chain molecules.
Figure~\ref{fig:scatter_OMC_857GHz} shows the Planck 857 GHz (350~$\micron$) map of the Orion A and B clouds, with the target positions overlaid.
White and magenta markers indicate sources with and without c-C$_3$HD detections, respectively. 
The targets with weak or no detections of c-C$_3$HD appear to be located near regions of bright 857 GHz emission ($F_\nu > 10^{-3.5}$ MJy sr$^{-1}$, see Figure~\ref{fig:scatter_OMC_857GHz}).
Since strong 857~GHz emission traces high dust column density and, consequently, high visual extinction, this suggests a possible correlation between dust extinction and the reduced abundance of carbon-chain molecules.

\begin{figure*}[htb!]
\centering
\includegraphics[width=1\linewidth]{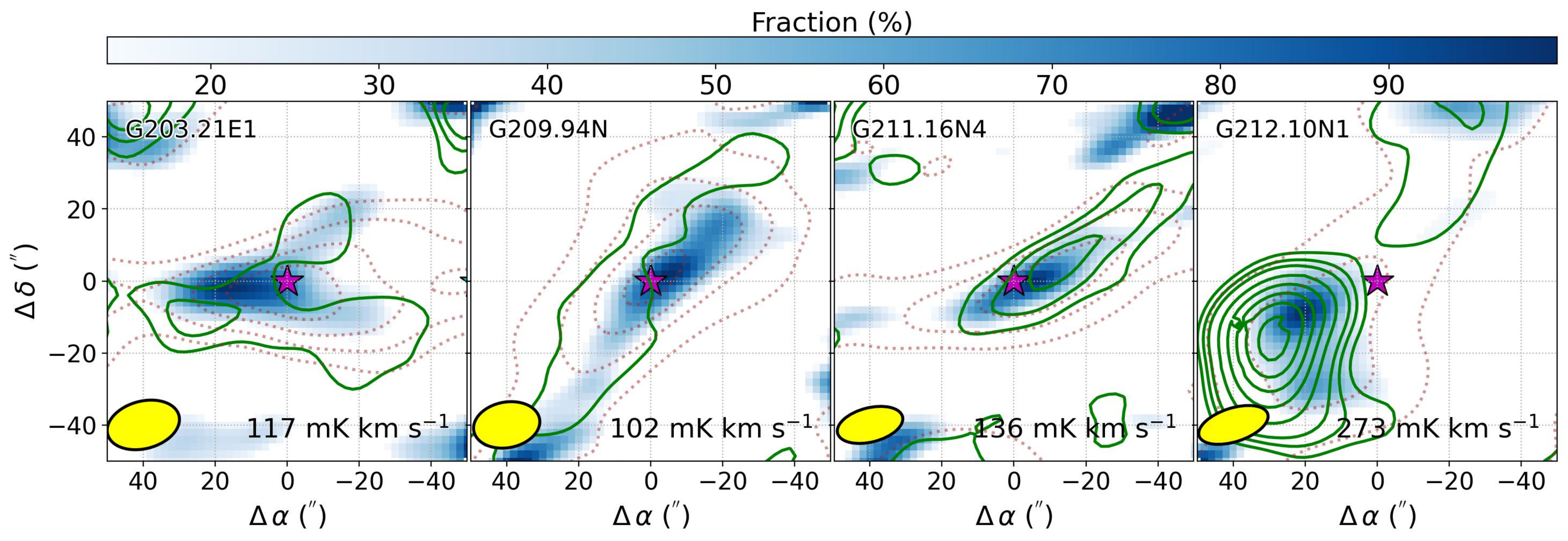}
\caption{\label{fig:mom0_CC} The integrated intensity images of c-C$_3$HD (rasters) overlaid with that of N$_2$H$^+$ (brown dotted contours) and CCS (green solid contours). 
The velocity range for deriving the integrated intensity images was fixed to $v_\mathrm{LSR}\pm1$ km s$^{-1}$. 
The yellow ellipse at the lower left corner of each panel shows the synthesized beam of the rasters achieved by the ACA. 
The text at the lower right corner of each panel corresponds the peak value for the global color bar.
For G209.94N and G212.10N1, the N$_2$H$^+$ contour levels are set at [5, 55, 105, 155, 205]$\sigma$; for the remaining two cores, the contour levels are [5, 25, 45, 65, 85]$\sigma$. 
The CCS contours are plotted at levels of [5, 10, 15, 20, 25]$\sigma$.
$\sigma\sim7.5$ mK km s$^{-1}$.
The magenta markers indicate the sources included in this study, with the target coordinate as the reference; the corresponding coordinates are listed in Table~\ref{tab:coord}.
The cyan markers indicate nearby sources reported by literature, which can be found in the overview description of each source in Appendix~\ref{appx:overview:mom0}. 
}
\end{figure*}

\subsection{Morphology}

To access the morphology, we generated integrated intensity (moment 0) maps for the four molecular species.
For each molecule of each source, a consistent velocity integration interval was applied across the spatial plane.
To determine this interval, we initially used the datacubes without primary beam correction.
We then extracted sub-cubes covering $\pm 3$ km s$^{-1}$ around the source $v_\mathrm{LSR}$, within a 25\arcsec\ (10,000 au) circular region centered on the target position.
While the SNR for determining the molecular detection was five for a more robust criteria, integration interval was defined as the longest contiguous range of velocity channels in which the signal exceeded three times the noise level.
The width of the integration interval is listed in Table~\ref{tab:appx:Ntot}. 
We finally applied the primary beam correction on the integrated maps. 
We include the resulting images derived from the aforementioned methods in Appendix~\ref{appx:overview:mom0} for completeness (Figures~\ref{fig:appx:mom0_G203.21E1} -- \ref{fig:appx:mom0_G212.10N1}). 
In the main text, we present Figures~\ref{fig:mom0_main} and \ref{fig:mom0_CC}, where a fixed integration interval of $v_\mathrm{LSR}\pm1$ km s$^{-1}$ is adopted. 
The overall morphologies are consistent with those derived from the noise-based intervals.

As shown in Figure~\ref{fig:mom0_main}, N$_2$H$^+$ is known to trace cold and dense gas because CO, which would otherwise destroy N$_2$H$^+$ in the gas phase, becomes depleted in these environments \citep[e.g., ][]{2002Bacmann_COdepletion}.
Most targets exhibit filamentary structures with a core near the center of the field, which we define as the primary starless core in this study.
Some targets, such as G205.46M3, G206.21N, G208.68N2, and G209.29S2, show clearly identifiable multiple N$_2$H$^+$ components within the field.
Some of these companion cores are newly detected in our study, while others were previously identified as starless or protostellar cores by surveys such as \citet{2018Yi_PGCC_SCUBA2_II} and \citet{2020Dutta_ALMASOP}.
Conversely, some known nearby YSOs are faint in radio emission and do not show corresponding N$_2$H$^+$ components in our observations.

The spatial distribution of CH$_3$OH emission exhibits diverse and fragmented morphologies.
It appears as core-like (e.g., G206.21S, G209.29N1, G209.55N2, G209.77E3, G210.37N); elongated and located on one side of the N$_2$H$^+$ component (G203.21E1, G209.94N, G211.64N4); elongated but aligned with N$_2$H$^+$ (G206.21N); extended (G209.29S1, G209.29S2, G210.82N2, G211.16N3); and complex (G205.46M3, G208.68N2, G212.10N1).
In general, CH$_3$OH is distributed at the periphery of N$_2$H$^+$, which was also observed in other  cores \citep[e.g., ][]{2006Tafalla_L1498_1517B, 2014Bizzocchi_L1544_CH3OH, 2014Vastel_L1544_COM, 2016Jimenez_L1544_COMs, 2018Ohashi_TUKH122, 2020Harju_H-MM1_CH3OH, 2025Hsu_ALMASOP_starless}. 
Based on high–spatial- and spectral–resolution observations, \citet{2025Hsu_ALMASOP_starless} proposed that CH$_3$OH traces turbulence-driven shocks of diffuse gas impinging onto dense filaments highlighted by N$_2$H$^+$.

\begin{figure*}[htb!]
\centering
\includegraphics[width=.9\textwidth]{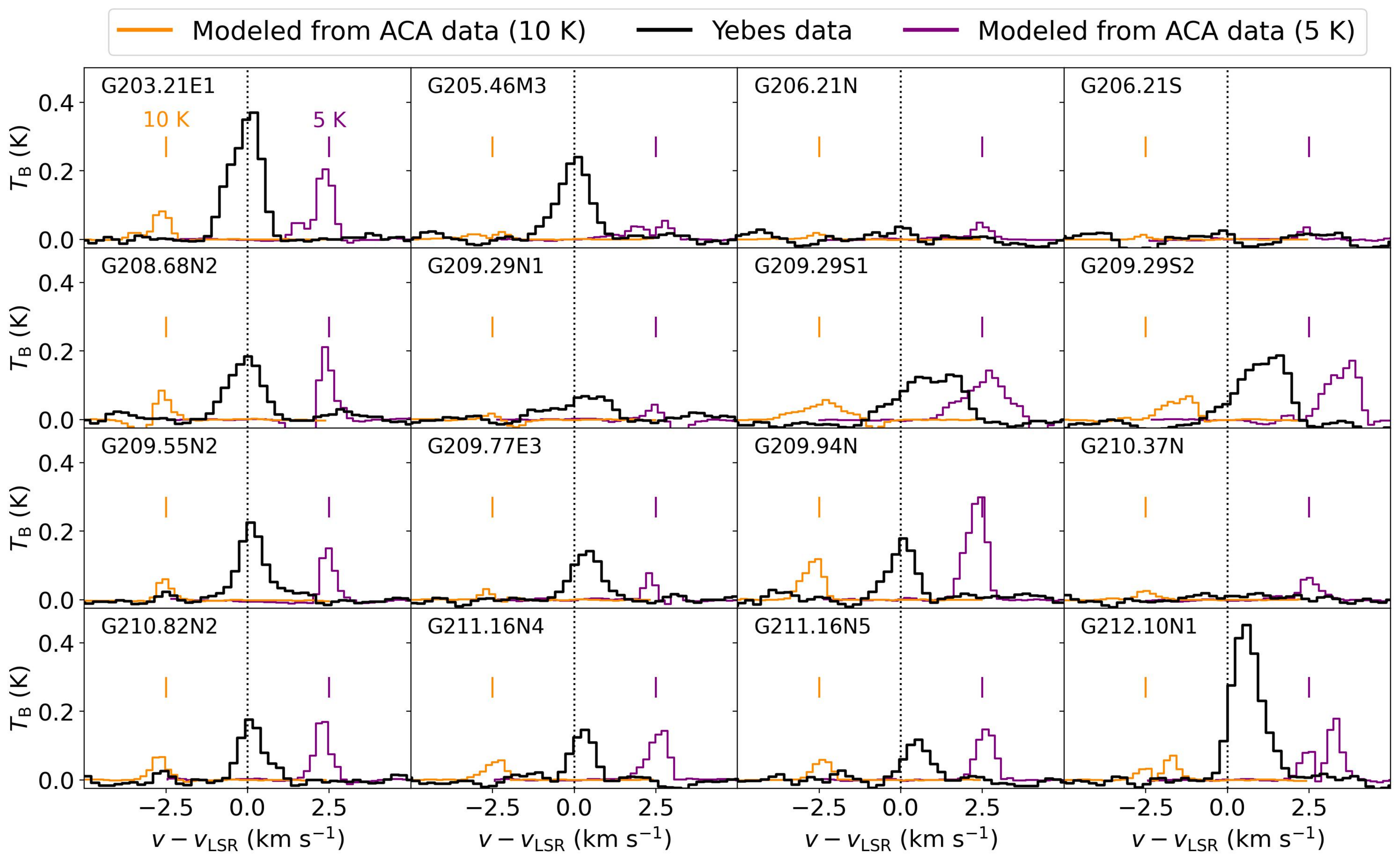}
\caption{\label{fig:spec_Yebes_ACA} 
The spectra observed with Yebes 40-m observations (black) and modeled from ACA observations (orange and purple) for the CH$_3$OH-3K transition. 
The orange and purple spectra were estimated with given rotational temperature at 10 and 5 K, respectively, with a velocity offset of -2.5 and 2.5 km s$^{-1}$, respectively. 
We note that the angular separation between G209.29S1 and G209.29S2 is comparable to the Yebes beam size (36\arcs), indicating that their emissions may be mutually contaminated. 
}
\end{figure*}

Both CCS and c-C$_3$HD are carbon-chain molecules, and in the six targets both CCS and c-C$_3$HD are consistently detected (Figure~\ref{fig:scatter_detection}).
The two carbon-chain molecules and CH$_3$OH generally have different distributions, as shown in the figures of Appendix~\ref{appx:overview:mom0}. 
Even though they are often detected simultaneously, CCS and c-C$_3$HD generally show distinct spatial distributions. 
In Figure~\ref{fig:mom0_CC}, we show the integrated intensity images of N$_2$H$^+$, c-C$_3$HD, and CCS with a fixed velocity range of $v_\mathrm{LSR} \pm 1$ km s$^{-1}$ for the selected sources.
These sources were selected due to their simultaneous detection of c-C$_3$HD, and CCS and without any nearby protostellar core. 
As shown in \ref{fig:mom0_CC} ( and corresponding panels in Figures \ref{fig:appx:mom0_G203.21E1}, \ref{fig:appx:mom0_G209.94N}, \ref{fig:appx:mom0_G211.16N4}, and \ref{fig:appx:mom0_G212.10N1}), both CCS and c-C$_3$HD are located in the outer layers surrounding the dense gas component traced by N$_2$H$^+$.
The misaligned peaks between CCS and N$_2$H$^+$ in four of our targets, G203.21E1, G298.68N2, G211.16N5, and G212.10N1, were also reported by \citet{2021Tatematsu_SCOPE}. 
In addition to the segregation between CCS and N$_2$H$^+$, as shown in Figure~\ref{fig:mom0_CC}, CCS appears to trace a more extended outer layer than c-C$_3$HD, which is particularly evident in the first two sources (G203.21E1 and G209.94N).

Despite the limited number of detected transitions for each molecule, we estimated the column density associated with each core.
Assuming optically thin and local thermodynamic equilibrium (LTE) with a fixed rotational temperature of 10~K \citep[a typical gas temperature of starless cores, e.g., ][]{2024Scibelli_COM_Perseus}, we derive the molecular column density maps for each species.
For the CH$_3$OH–7K transition, \citet{2024Scibelli_COM_Perseus} reported optical thickness ($\tau > 1$) in one starless core.
In our sample, the line peak brightness temperatures are all below or comparable to 1~K, which is significantly lower than the expected kinetic temperature of 10~K.
On the basis that CH$_3$OH emission is extended, the comparably low brightness temperature indicates that the emission in our sources is likely optically thin.
The collisional rate coefficient of this transition at 10~K is $7.6\times10^{-11}$~cm$^{3}$~s$^{-1}$ \citep{LAMDA:2005}, corresponding to a critical density of $4.5\times10^{4}$~cm$^{-3}$.
It is lower than or comparable to the typical gas densities in starless cores ($\sim10^{4}$–$10^{5}$~cm$^{-3}$), supporting the assumption of LTE.
For N$_2$H$^+$, the flux was multiplied by a factor of nine, as the $J=1-0$ transition components ratio should be 1:5:3 for the three hyperfine transition groups \citep{2006Daniel_N2Hp_HfS}. 

Appendix~\ref{appx:overview:mom0} presents the equations and Figures~\ref{fig:appx:mom0_G203.21E1} -- \ref{fig:appx:mom0_G212.10N1} for the results. 
Moreover, we list in Table~\ref{tab:appx:Ntot} the peak column densities for each species within a circular aperture of 25\arcs (10,000~au) centered on each target position.
Note that the density of the column $N^{p}_\mathrm{tot}$ may be overestimated due to nearby bright sources within the 25\arcs aperture.


\section{Comparisons with Single-dish Observations}
\label{sec:results:Yebes}

As described in the previous section, CH$_3$OH generally appears to be spatially extended.
The flux of such extended emission is likely underestimated due to interferometric spatial filtering.
To assess the amount of missing flux, we compare our data with observations from the Yebes 40-m telescope.
The strongest CH$_3$OH transition covered by the observation was the CH$_3$OH-3K transition. 

Figure~\ref{fig:spec_Yebes_ACA} shows the CH$_3$OH spectra (black) obtained from the Yebes 40-m observations.
Despite that we detected CH$_3$OH toward all the sources with the ACA observations, we did not see significant detection (signal-to-noise ratio, SNR $\geq$ 5) in G206.21N, G206.21S, G209.29N1, and G210.37N in Yebes observations. 

To assess the missing flux, we compared the CH$_3$OH lines observed with the Yebes 40-m single-dish telescope to the ACA interferometric observations.
To do so, we modeled the CH$_3$OH-3K spectrum based on the directly observed CH$_3$OH-7K spectrum.
First, we used the \texttt{imsmooth} task in \texttt{CASA} to convolve the ACA datacube to a circular beam comparable to the Yebes 40-m aperture (FWHM = 36\arcsec\ at 48.4~GHz).
We then extracted the CH$_3$OH-7K transition spectra at the pixel corresponding to the position targeted by the Yebes observations.
Finally, we calculated the expected CH$_3$OH-3K spectra by rescaling the observed CH$_3$OH-7K spectra with the equations in Appendix \ref{appx:overview:mom0}, assuming optically thin emission and LTE. 

The modeled spectra are also shown in Figure~\ref{fig:spec_Yebes_ACA}.
The orange and purple profiles represent the modeled spectra assuming rotation temperatures ($T_\mathrm{rot}$) of 10 K and 5 K, respectively, with velocity offsets of $-2.5$ and $+2.5$ km s$^{-1}$, respectively. 
A higher assumed rotational temperature results in a weaker modeled spectrum.
In general, the modeled spectra are weaker than those observed with the Yebes 40-m telescope, regardless of whether $T_\mathrm{rot}$ is 10 K or 5 K.
This difference suggest that the missing flux is significant in the ACA observations. 
In G206.21N, G206.21S, and G210.37N, the modeled spectra fall below the Yebes noise level, consistent with the non-detections in the Yebes observations.

\begin{figure}[htb!]
\centering
\includegraphics[width=.9\linewidth]{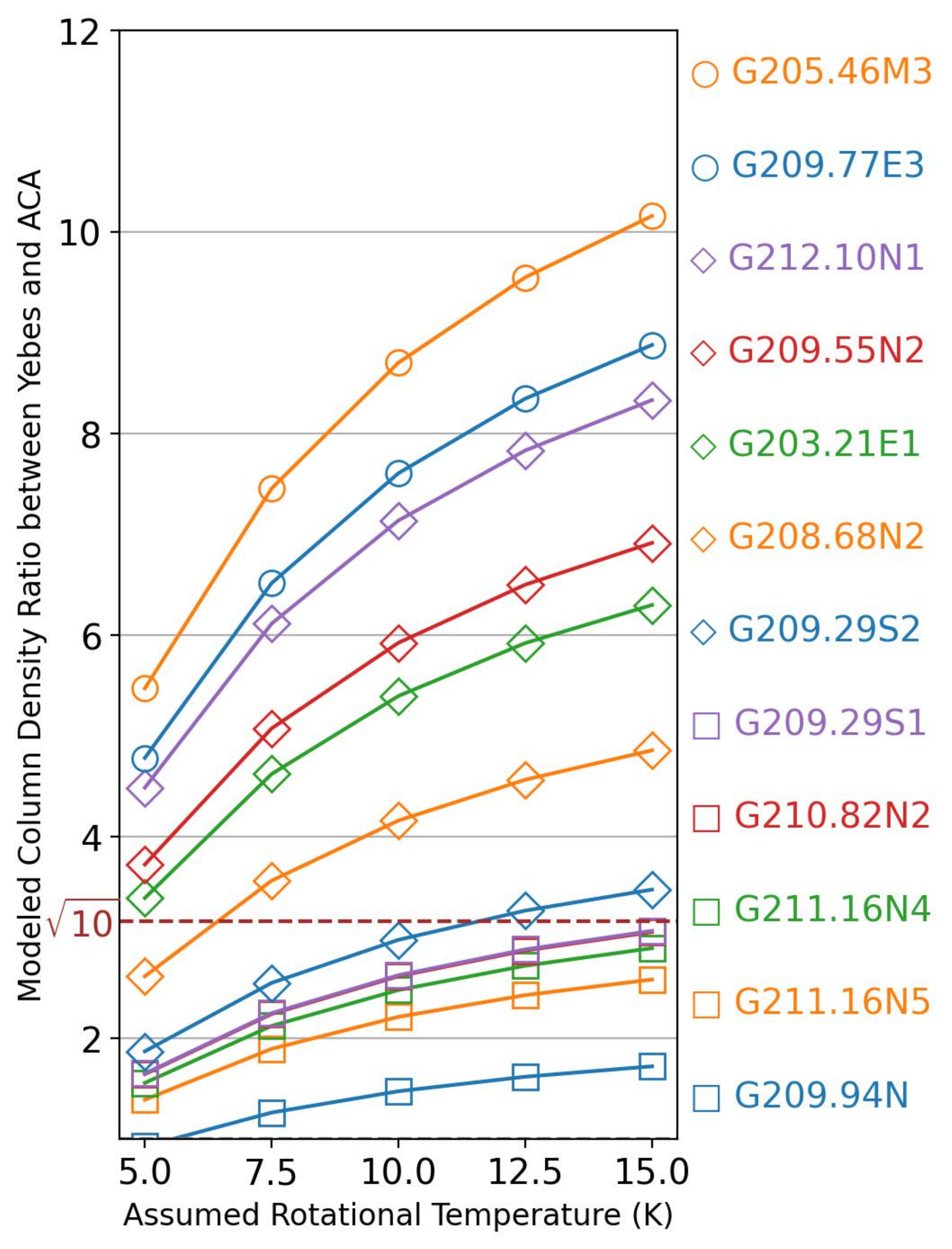}
\caption{\label{fig:Ntot_Yebes_ACA} Variation of inferred CH$_3$OH column density ratios between Yebes and ACA observations as a function of assumed rotational temperatures. 
The Yebes and ACA values were estimated from the observed spectra of the CH$_3$OH-3K and CH$_3$OH-7K transitions, respectively. 
Each polyline represents one source.
Only the 12 sources with CH$_3$OH detections in the Yebes observations are included in this plot. 
We note that the angular separation between G209.29S1 and G209.29S2 (36\arcs) is smaller than the Yebes beam size (36\arcs), indicating that their emissions may be mutually contaminated. 
The dashed line indicates a difference of one order of magnitude ($\sqrt{10}$). 
}
\end{figure}

To quantitatively evaluate the flux filtered out by the interferometer, we calculated the CH$_3$OH column densities at a common beam derived from the ACA and Yebes observations, with the equations in Appendix~\ref{appx:Obs} under LTE, based on the observed CH$_3$OH-7K and CH$_3$OH-3K transitions, respectively.
The results are shown in Figure~\ref{fig:Ntot_Yebes_ACA}.
As illustrated in the figure, the ratio of the column density derived from the Yebes observations to that from the ACA data increases with increasing assumed rotational temperature. 

At $T_\mathrm{rot} = 10$ K, a typical gas temperature of starless cores \citep[e.g.,][]{2024Scibelli_COM_Perseus}, the observed column densities of the Yebes data can be up to ten times of those derived from the ACA observations.
\citet{2020Scibelli_COM_Taurus} investigated CH$_3$OH in starless cores within the Taurus cloud and found that the typical CH$_3$OH gas kinetic temperature is about 7.5~K.
At this temperature, six sources in our sample show column density ratios greater than $\sqrt{10}$, indicating that the ACA observations may be missing an order of magnitude of CH$_3$OH flux. 
These results highlight the significant amount of missing flux when observing CH$_3$OH emission with interferometry alone.

\section{Discussions}
\label{sec:Disc}

\subsection{Chemical segregation related to CH$_3$OH}
\label{sec:Disc:seg}

\subsubsection{CH$_3$OH vs. N$_2$H$^+$}
\label{sec:Disc:seg:N2Hp}
Our ACA observations show the general mismatched peak positions among N$_2$H$^+$, CH$_3$OH, c-C$_3$HD, and CCS. 
On top of that, N$_2$H$^+$ is tracing the cold and dense core, while the other species seemingly reside in the outer layer of the core. 
The shell-like morphology of CH$_3$OH surrounding the cold gas tracers (e.g., N$_2$H$^+$) or dust in starless  cores were also reported in the literature. 
For example, \citet{2006Tafalla_L1498_1517B} found that, in L1498 and L1517B, molecules such as CH$_3$OH and c-C$_3$H$_2$ fit with profiles having a sharp central hole. 
\citet{2014Bizzocchi_L1544_CH3OH} reported that in the prestellar core L1544, the CH$_3$OH emission resembles a non-uniform ring surrounding the dust peak. 
\citet{2014Vastel_L1544_COM} applied the non-LTE analysis of the methanol lines in L1544 and shows that they are likely emitted at the border of the core at a radius of $\sim$ 8,000 au. 
\citet{2016Jimenez_L1544_COMs} compare the molecular composition between the continuum peak and the CH$_3$OH peak in L1544 and found that the COMs at the CH$_3$OH peak are overall more abundant that that at the continuum peak. 
\citet{2016Jimenez_L1544_COMs} proposed a scenario to explain the shell-like morphology of CH$_3$OH. 
In this scenario, three conditions must be met:
(1) the density must be high enough for efficient CO freeze-out, allowing CH$_3$OH to form on grain surfaces;
(2) the density must not be too high, so that gaseous CH$_3$OH is not significantly depleted; and
(3) the visual extinction must be sufficiently large to shield CH$_3$OH from photodissociation by the external interstellar radiation field (ISRF).
Together, these conditions suggest that gaseous CH$_3$OH can only survive within a shell-like layer at intermediate densities. 


\subsubsection{CH$_3$OH vs. carbon-chains molecules}
\label{sec:Disc:seg:CC}

Anti-correlations between CH$_3$OH and the two carbon-chain molecules (CCS and c-C$_3$HD) are observed in our studies (corresponding panels in Figures \ref{fig:appx:mom0_G203.21E1}--\ref{fig:appx:mom0_G212.10N1}). 
Chemical segregation between CH$_3$OH and c-C$_3$H$_2$ has been reported in well-studied cores such as L1544 \citep{2016Spezzano_L1544_chem, 2017Spezzano_L1544_chem}.
\citet{2016Spezzano_L1544_chem} suggested that such segregation between saturated (CH$_3$OH) and unsaturated (carbon chain) organics is driven by variations in the local ISRF.
Regions exposed to stronger ISRF maintain a higher abundance of atomic carbon in the gas phase, favoring the formation of carbon-chain molecules such as c-C$_3$H$_2$.
In contrast, in more shielded regions, carbon is predominantly locked in CO, allowing the efficient formation of COMs such as CH$_3$OH on dust grain surfaces.
The illumination effect was also supported by \citet{2020Spezzano_survey_chem}, who investigated the column density ratio between CH$_3$OH and c-C$_3$H$_2$, $N$(CH$_3$OH)/$N$(c-C$_3$H$_2$), in six starless cores embedded in environments with varying levels of shielding and at different evolutionary stages.
For each target, the authors defined the ``head'' and ``tail'' based on the steep and shallow H$_2$ column density gradients, respectively. They found that the CH$_3$OH-to-c-C$_3$H$_2$ column density ratio is generally lower in the ``head'' than in the ``tail,'' suggesting that the chemical segregation is driven by uneven illumination.


\subsection{The Ubiquitous CH$_3$OH Gas in Starless Cores}
\label{sec:CH3OH:ubiquity}

The presence of gaseous CH$_3$OH in star-forming regions is generally attributed to its formation on dust grain surfaces, followed by desorption into the gas phase \citep{2009Herbst_COM_review}, despite other pathways were also suggested \citep[e.g., reaction of CH$_3^+$ with water ice, ][]{2023Nakai_CH3OH_formation}. 

In the protostellar stage, gas-phase COMs are primarily produced through thermal desorption.
As the temperature rises, COM molecules can be partially released from ice surfaces into the gas phase, and are fully desorbed when the temperature exceeds the ice sublimation threshold of $\sim$100~K.
A major heating source is the central protostar, which warms the inner envelope and creates a warm region rich in gaseous COMs, commonly referred to as a hot corino \citep[e.g.,][]{2004Ceccarelli_HotCorino,2023Hsu_ALMASOP}.
Non-thermal desorption processes also play an important role.
One such mechanism in protostellar sources is ice sputtering caused by energetic shocks.
These shocks can originate from outflows \citep[e.g.,][]{2008Arce_L1157-B1_COMs,2024Hsu_HOPS87,2025Bouvier_HOPS409} or from envelope-to-disk accretion \citep[e.g., ][]{2002Velusamy_L1157_CH3OH, 2016Oya_IRAS16293-2422, 2022Okoda_B335_chem,2025Hsu_G192}.
Shocks can also locally raise temperatures and trigger thermal desorption. 

In the starless stage (including the prestellar stage), however, the absence of a central protostar, outflows, envelope, and disk structure makes the above-mentioned desorption scenarios unlikely and suggests the need for different mechanisms. 
These reactions, such as 
reactive desorption \citep[e.g., ][]{2007Garrod_reactive-desorption,2013Vasyunin_reactive-desorption,2017Vasyunin_reactive-desorption,2018Chuang_reactive-desorption}, 
cosmic-ray-induced sputtering \citep{2020Dartois_CR-sputtering,2021Wakelam_CR_sputtering}, 
and shocks \citep[e.g., ][]{2015Soma_TMC1, 2020Harju_H-MM1_CH3OH, 2020Scibelli_COM_Taurus, 2022Lin_L1544_CH3OH_HNCO, 2022Kalvans_desorption, 2025Hsu_ALMASOP_starless}, can lead to the desorption of either COMs or their precursors. 
In reactive desorption, the desorption occurs naturally as a result of the exothermic reactions , such as those involved in the formation of methanol on grains. 
Cosmic-ray-induced sputtering occurs when cosmic rays impact solids, and the excitation of electrons leads to a brief thermal spike that can eject surface species.
In addition to the two mechanisms mentioned above, the studies by \citet{2007Roberts_desorption} and \citet{2025Priestley_COM_MHD} both addressed three other desorption mechanisms, H$_2$ formation on grains, direct cosmic-ray heating, and cosmic-ray–induced photodesorption.

Shocks may be caused by turbulent activity or mass accretion flows.
In TMC-1 CP, \citet{2015Soma_TMC1} proposed the clump–clump collision shock scenario as a possible mechanism for CH$_3$OH desorption.
In H-MM1, \citet{2020Harju_H-MM1_CH3OH} suggested that CH$_3$OH is released as a result of grain–grain collisions induced by turbulence.
\citet{2020Scibelli_COM_Taurus} found that the CH$_3$OH line profiles in their targeted starless cores could be explained by a combination of unresolved motions and supersonic turbulence.
\citet{2022Lin_L1544_CH3OH_HNCO} proposed that the enhancement of CH$_3$OH in L1544 is driven by slow shocks induced by accretion flows linked to the large-scale cloud dynamics.
In this scenario, the extent of CH$_3$OH is primarily determined by the location of shocks, while environmental effects, such as the external illumination \citep{2016Spezzano_L1544_chem,2020Spezzano_survey_chem}, are not necessarily in conflict with this interpretation. 
On another note, \citet{2025Spezzano_IRAS16293E_COM} detected red- and blue-shifted components in addition to the primary velocity component of CH$_3$OH toward IRAS 16293E. 
They proposed that these shifted components do not originate from the core itself but instead arise from line-of-sight emission produced by shocks, likely associated with outflows from a nearby protostar system. 
This interpretation was subsequently supported by \citet{2025Scibelli_IRAS16293E_COM}.

Recently, under our ALMASOP project, \citet{2025Hsu_ALMASOP_starless} identified shock interfaces between N$_2$H$^+$ and CH$_3$OH in several starless cores and proposed a ``turbulence-induced mass-assembly shock'' scenario.
In this picture, magnetohydrodynamic (MHD) turbulence facilitates mass assembly in starless cores \citep[][]{2001Balsara_turbulent-MHD-model,2016Ohashi_TUKH122_turbulence,2017Keown_CepheusL1251}, while the accompanying shocks drive CH$_3$OH desorption from icy grain mantles.
The desorption may result either from localized temperature increases (i.e., thermal desorption) \citep[e.g.,][]{2001Dickens_TMC1,2015Soma_TMC1} or from mechanical processes such as sputtering or grain disruption (i.e., non-thermal desorption) \citep[e.g.,][]{2022Kalvans_desorption}.

In our ACA observations, all the targeted starless cores exhibit detectable CH$_3$OH emission, even though four of them are not detected in the Yebes observations. 
This discrepancy likely arises from the larger beam size of the Yebes observations ($\sim$36\arcs) compared to the ACA observations ($\sim$16\arcs), despite their similar noise levels in brightness temperature ($\sim$10~mK). 
The ubiquity of CH$_3$OH in starless cores has also been reported in surveys toward the Taurus \citep{2020Scibelli_COM_Taurus} and Perseus \citep{2024Scibelli_COM_Perseus} molecular clouds. 
As all the three surveys (including this study) focus on low-/intermediate-mass cores, the ubiquity of CH$_3$OH suggests that desorption processes, whether thermal or non-thermal, are prevalent in these starless cores.
Given the uniformly high detection rates under diverse physical conditions, detection statistics alone do not allow any desorption mechanism to be ruled out.

Although CH$_3$OH is commonly distributed near the surface of dense cores traced by N$_2$H$^+$, our maps reveal that the former in general does not fully cover the latter (Figure~\ref{fig:mom0_main}).
In other words, CH$_3$OH sometimes traces only one side of the filament (e.g., G203.21E1 and G209.29S1/S2) or appears as a localized spot near its surface.
Such asymmetric distributions have often been attributed to uneven exposure to the ISRF, as discussed in Sect.~\ref{sec:Disc:seg:CC}.
A potential indication of this effect in our study is the common asymmetric CH$_3$OH distribution in the five southern cores, G210.37N, G210.82N2, G211.16N4/N5, and G212.10N1.
These five cores lie within four clumps that exhibit comparable large-scale dust distributions (the four white markers in Figure~\ref{fig:scatter_OMC_857GHz}) and their CH$_3$OH emission generally appears on the eastern side of the N$_2$H$^+$.
This common asymmetry might result from stronger ISRF on the general western side of the clumps where the cores are embedded within, although we fail to identify promising stellar sources responsible for such enhanced (UV) radiation fields.

Under the turbulence-induced mass-assembly shock scenario, such asymmetric CH$_3$OH distributions reflect local turbulence activity.
As described in Sect.~\ref{sec:results:ACA}, CH$_3$OH appears to trace diffuse gas components near the filament surface itself. 
This scenario naturally explains why CH$_3$OH exhibits substructures that are not seen in N$_2$H$^+$.
In this framework, the localized CH$_3$OH spots likely trace shock fronts at the interfaces between the diffuse and dense gas components, where mass flows converge.

\subsection{The Extended CH$_3$OH Component inferred from the Missing Flux}
\label{sec:CH3OH:extended}

Our comparison between the Yebes and ACA observations suggests that the ACA data can significantly underestimate the emitted CH$_3$OH flux. 
This underestimation points to the presence of a spatially extended component, potentially larger than 60\arcsec (the maximum recoverable scale of the ACA). 
Images from single-dish observations are necessary to confirm this. 
We searched the literature for supporting evidence of such an extended component and found that, in L1544, the CH$_3$OH morphology indeed shows a superposition of an enhanced component on the eastern side of the dense core overlaid on a more extended emission structure.
This is illustrated in Figure 1 of \citet{2017Spezzano_L1544_chem} and \citet{2022Lin_L1544_CH3OH_HNCO}, both based on observations with the IRAM 30-m single-dish telescope. 
Since L1544 is a prestellar core, which is more evolved among starless cores, more sensitive single-dish surveys toward starless cores, including both prestellar and non-prestellar ones, are needed to further confirm this result.

A similar case was reported by \citet{2020Scibelli_COM_Taurus}, who performed on-the-fly (OTF) mapping over $15\arcmin\times15\arcmin$ regions containing starless cores in the Taurus molecular cloud.
They detected methanol emission extending up to $\sim$15\arcmin\ ($\sim$1.2$\times$10$^5$ au) in two regions, B10 and B211, which were classified as ``less evolved'' based on the absence of protostellar activity \citep{2015Seo_TMC}.
Furthermore, \citet{2020Scibelli_COM_Taurus} found that starless cores within these less evolved regions tend to exhibit higher CH$_3$OH fractional abundances.
They interpreted this trend as a consequence of depletion: in more evolved regions, cores are generally denser, leading to larger CH$_3$OH depletion zones. 

It is intriguing that CH$_3$OH gas is detected in such diffuse regions.
One possible explanation is that CH$_3$OH forms in situ on dust grain surfaces and subsequently desorbs.
Alternatively, CH$_3$OH molecules could form and desorb in denser regions and then be transported into the surrounding diffuse gas.
Both cases could be explained by the turbulence-induced shock scenario.
In the former, CH$_3$OH may trace ongoing turbulent activity that could drive non-thermal desorption.
In the latter, the emission might be related to the mixing between the cloud gas and the surrounding medium due to the instabilities near the surface with strong velocity shear \citep[e.g.,][]{2021Kupilas_model}.
Future simulations that include both turbulent dynamics and the formation, desorption, and adsorption processes of CH$_3$OH may help to assess this possibility.


\subsection{Segregation between Carbon-chain Molecules}
\label{sec:Disc:CC}

Carbon-chain molecules can be efficiently produced in starless cores \citep[e.g.,][and the references therein]{2013Sakai_review_carbon-chain}.
In the early diffuse-cloud stage, C$^+$ is the dominant form of carbon in the gas phase because interstellar UV radiation penetrates deeply.
C$^+$ can then participate in a series of gas-phase reactions with species such as H$_2$, e$^-$, H$_3^+$, and C to form abundant carbon-chain molecules \citep[see Figure 1 in][]{2013Sakai_review_carbon-chain}.
As the extinction increases, carbon becomes locked in the form of CO, and carbon-chain chemistry is suppressed. 

In our observations, the two carbon-chain molecules, c-C$_3$HD and CCS, are commonly detected together.
Moreover, starless cores within the same clump generally show similar patterns of detection or non-detection for carbon-chain molecules.
Clumps located in regions bright at 857~GHz generally appear deficient in carbon-chain chemistry, suggesting an anti-correlation between extinction and the presence of carbon-chain species.
These trends are consistent with the current understanding of the chemical network of carbon-chain molecules, in which higher extinction leads to more carbon being locked in CO, thereby suppressing carbon-chain chemistry \citep[e.g., ][and the references therein]{2013Sakai_review_carbon-chain,2024Taniguchi_review}.

Interestingly, our observations show that although c-C$_3$HD and CCS are frequently co-detected in the same sources, their emission exhibit different distributions.
As described in Section~\ref{sec:results:ACA} and shown in Figure~\ref{fig:mom0_CC}, CCS appears to trace a more extended outer layer than c-C$_3$HD, which may reflect that c-C$_3$H$_2$ and CCS generally do not reside in the same structures. 
This contrasts with \citet{2017Spezzano_L1544_chem}, who found that in the prestellar core L1544, c-C$_3$H$_2$ (the main isotopologue) and CCS share the same emission peak.
Admittedly, spatial filtering by interferometers may affect the observed morphology.
The more centrally concentrated distribution of c-C$_3$HD might also be explained by deuterium enhancement in dense, cold environments \citep[e.g.,][]{2001Rodgers_deuterium,2022Giers_L1544_cC3H2}.
Nevertheless, non-coincident distributions between c-C$_3$H$_2$ and CCS have been observed toward other starless cores through single-dish imaging.
For example, based on the images presented by \citet{2020Spezzano_survey_chem} with the IRAM 30-m telescope, such non-coincident distributions are seen in Oph D, HMM1, and L694-2, and possibly in L429 as well, although this was not explicitly discussed in the text.
Taken together, these results suggest that c-C$_3$H$_2$ (and c-C$_3$HD) and CCS generally exhibit distinct spatial extents.

To further examine whether the layered onion-shell like distribution of CCS and c-C$_3$HD is due to excitation effects, we estimated their critical densities.
As shown in Table~\ref{tab:molec_info}, CCS and c-C$_3$HD have comparable Einstein coefficients. 
The collision rate coefficient ($\gamma_{ij}$) of the CCS transition at 10~K is $\sim2\times10^{-12}$ cm$^{3}$ s$^{-1}$ in the LAMDA \footnote{\url{https://home.strw.leidenuniv.nl/~moldata/}} \citep{LAMDA:2005} database.
Meanwhile, given the lack of c-C$_3$HD collision rate in LAMDA, we adopted the value of c-C$_3$H$_2$ of the same transition ($\sim9\times10^{-12}$ cm$^{3}$ s$^{-1}$), which is significantly higher than that of CCS.
Consequently, $c$-C$_3$HD emission has a lower critical density than CCS emission and can trace more diffuse gas, which can not explain the observed morphology.

The layered distributions of CCS and c-C$_3$H$_2$ (and c-C$_3$HD) may arise from carbon-chain chemistry and the poorly understood sulfur chemistry. 
One possible explanation is the photo-dissociation of carbon-chain molecules, under the framework of photo-chemistry, 
Irradiation can prevent carbon from being efficiently retained in CO molecules, thereby enhancing carbon-chain chemistry. However, in regions most strongly exposed to irradiation, carbon-chain species themselves may be dissociated, ultimately suppressing the formation of larger chains.
As a result, CCS becomes relatively abundant in the outer layers, while deeper layers can harbor more complex carbon-chain molecules such as c-C$_3$H$_2$ (and thus c-C$_3$HD).
Within this framework, even larger carbon-chain molecules such as CCCS would be expected to show distributions more similar to c-C$_3$H$_2$.

Alternatively, CCS abundance in the outer layers may be enhanced through chemical processing of c-C$_3$H$_2$ (or c-C$_3$HD).
Photo-chemistry could facilitate this pathway by producing enhanced S$^+$ in the outer layers, which in turn promotes the formation of CCS.
According to the reactions in KIDA\footnote{\url{https://kida.astrochem-tools.org/}} \citep[KInetic Database for Astrochemistry, ][]{KIDA:2012, KIDA:2024} and Fig.~1 in \citet{2024Taniguchi_review}, CCS can form from c-C$_3$H$_2$ via:
\begin{equation}
    c{\text -}\mathrm{C_3H_2+S^+ \rightarrow HCCCS^+ +H}
\end{equation}
\begin{equation}
    \mathrm{HCCCS^+ + e^- \rightarrow CCS + CH}.
\end{equation} 
HCCCS$^+$ is also a parent species of CCCS:
\begin{equation}
    \mathrm{HCCCS^+ + e^- \rightarrow CCCS + H. }
\end{equation}
As a result, within this framework, CCS and CCCS should share similar distributions. 
High-spatial resolution images or the line profiles of c-C$_3$H$_2$, CCS, and CCCS from single-dish observations will help verify the scenarios.
Also, the density-based clustering method with a variety of features such as the intensity, velocity offset, and linewidth could possibly provide insights on the differentiation between molecular species, as demonstrated by \citet{2025Giers_chem_DBSCAN}.



\section{Conclusions}
\label{sec:Conclusions}

We conducted chemical surveys toward a sample of 16 starless cores in the Orion cloud selected from the ALMASOP catalog with ACA and Yebes observatories. 
We report the detection of N$_2$H$^+$, CH$_3$OH, CCS, and c-C$_3$HD. 

\begin{enumerate}

  \item We report, in ACA observations, 100\% detection rates for both N$_2$H$^+$ and CH$_3$OH, 62.5\% for CCS, and 50\% for c-C$_3$HD. This suggests that CH$_3$OH is prevalent in starless cores, consistent with other surveys of starless cores in the Taurus and Perseus clouds. 
  The ACA images reveal that CH$_3$OH and the two carbon-chain molecules generally trace the outer layers of the dense core traced by N$_2$H$^+$, each exhibiting distinct spatial distributions. 
  
  \item Comparisons between the ACA and Yebes observations suggest that the ACA data may underestimate the CH$_3$OH flux by up to an order of magnitude, depending on the assumed methanol rotational temperature.
  These findings indicate the presence of an extended and flattened CH$_3$OH gas component.

  \item We find that clumps located near dust-rich regions on larger scales tend to lack detections of carbon-chain molecules.
  This trend suggests that environmental effects at the clump scale, possibly linked to the strength of the interstellar radiation field, as discussed in the literature, can influence the organic chemical composition.
  In addition, we identify chemical segregation between the two carbon-chain species CCS and c-C$_3$HD within individual starless cores.
  These distinct morphologies may likewise be understood in the context of photo-chemistry, highlighting the role of irradiation in regulating carbon-chain chemistry across both clump and core scales.
  
\end{enumerate}


\acknowledgments
This paper makes use of the following ALMA data: ADS/JAO.ALMA\#2021.2.00094.S. ALMA is a partnership of ESO (representing its member states), NSF (USA), and NINS (Japan), together with NRC (Canada), NSTC and ASIAA (Taiwan), and KASI (Republic of Korea), in cooperation with the Republic of Chile. The Joint ALMA Observatory is operated by ESO, AUI/NRAO, and NAOJ.
Based on observations carried out with the Yebes 40 m telescope (22A010). The 40 m radio telescope at Yebes Observatory is operated by the Spanish Geographic Institute (IGN; Ministerio de Transportes y Movilidad Sostenible). 
X.-C. Liu and T. Liu acknowledges the supports by the National Key R\&D Program of China (No. 2022YFA1603101). 
S.-Y. Hsu acknowledges supports from the Academia Sinica of Taiwan (grant No. AS-PD-1142-M02-2) and National Science and Technology Council of Taiwan (grant No. 112-2112-M-001- 039-MY3).  
S.-Y. Liu acknowledges supports from National Science and Technology Council of Taiwan (grant No. 113-2112-M-001-004- and 114-2112-M-001-035-MY3).

\software{
\texttt{Astropy} \citep{astropy:2013, astropy:2018, astropy:2022},
\texttt{CASA} \citep{casa:2007,casa:2022},
\texttt{CARTA}  \citep{2021Comrie_CARTA}, 
\texttt{GILDAS} \citep{2005GILDAS}.
}

\clearpage
\appendix

\resetapptablenumbers
\section{Observational Parameters}
\label{appx:Obs}

Tables~\ref{tab:appx:spw}, \ref{tab:appx:obs} and \ref{tab:appx:noise} show the ACA observational beam sizes and noise levels, respectively, for the continuum and each spectral window (SPW).
In addition, the noise level of the Yebes observations is also listed in Table~\ref{tab:appx:noise}.

\begin{deluxetable}{rrrrrl}
\label{tab:appx:spw}
\caption{
Information of the spectral windows. 
}
\tablehead{
\colhead{SPW} & \colhead{$f_\mathrm{c}$} & \colhead{BW} & \colhead{d$f$} &\colhead{d$v$} & \colhead{Molecule} \\
\colhead{} & \colhead{(GHz)} & \colhead{(MHz)} & \colhead{(kHz)} &\colhead{(km s$^{-1}$)} & \colhead{}
}
\startdata
\#16 & 93.164 & 59 & 61 & 0.18 & N$_2$H$^+$ \\
\#18 & 93.863 & 59 & 61 & 0.18 & CCS \\
\#26 & 96.732 & 59 & 61 & 0.18 & CH$_3$OH \\
\#42 & 107.003 & 117 & 122 & 0.37 & c-C$_3$HD 
\enddata
\tablecomments{
$f_\mathrm{c}$ is the center frequency. 
BW is the bandwidth. 
d$f$ and d$v$ are the spectral resolution and the corresponding velocity resolution, respectively. 
}
\end{deluxetable}

\begin{deluxetable}{llcccccc}
\caption{\label{tab:appx:obs} Summary of the ACA observation beams.}
\tablehead{
\colhead{Name} & \colhead{Short Name} & \colhead{3~mm} & \colhead{N$_2$H$^+$} & \colhead{CCS} & \colhead{CH$_3$OH} & \colhead{c-C$_3$HD} \\
\colhead{} & \colhead{} & \colhead{(MAJ,MIN,PA)} & \colhead{(MAJ,MIN,PA)} & \colhead{(MAJ,MIN,PA)} & \colhead{(MAJ,MIN,PA)} & \colhead{(MAJ,MIN,PA)} 
}
\startdata
G203.21-11.20E1 & G203.21E1 & 15\farcs4, 12\farcs5, 76$^{\circ}$ & 19\farcs5, 15\farcs8, -77$^{\circ}$ & 16\farcs8, 13\farcs8, 76$^{\circ}$ & 22\farcs0, 14\farcs8, -74$^{\circ}$ & 20\farcs5, 13\farcs2, -75$^{\circ}$ \\
G205.46-14.56M3 & G205.46M3 & 15\farcs4, 12\farcs1, 80$^{\circ}$ & 18\farcs5, 15\farcs4, 73$^{\circ}$ & 18\farcs1, 15\farcs4, 76$^{\circ}$ & 17\farcs5, 15\farcs2, -73$^{\circ}$ & 16\farcs4, 13\farcs1, -80$^{\circ}$ \\
G206.21-16.17N  & G206.21N  & 15\farcs4, 12\farcs0, 80$^{\circ}$ & 18\farcs7, 15\farcs7, -74$^{\circ}$ & 17\farcs5, 15\farcs2, -83$^{\circ}$ & 17\farcs0, 14\farcs8, -83$^{\circ}$ & 16\farcs3, 13\farcs0, -81$^{\circ}$ \\
G206.21-16.17S  & G206.21S  & 15\farcs4, 12\farcs0, 82$^{\circ}$ & 18\farcs5, 15\farcs4, -75$^{\circ}$ & 17\farcs7, 15\farcs1, -81$^{\circ}$ & 17\farcs2, 14\farcs5, -83$^{\circ}$ & 16\farcs2, 12\farcs7, -81$^{\circ}$ \\
G208.68-19.20N2 & G208.68N2 & 19\farcs4, 9\farcs7, -73$^{\circ}$ & 21\farcs3, 14\farcs5, -79$^{\circ}$ & 21\farcs4, 14\farcs6, -77$^{\circ}$ & 20\farcs7, 14\farcs1, -77$^{\circ}$ & 18\farcs5, 12\farcs7, -78$^{\circ}$ \\
G209.29-19.65N1 & G209.29N1 & 19\farcs3, 9\farcs6, -73$^{\circ}$ & 21\farcs3, 14\farcs5, -80$^{\circ}$ & 21\farcs3, 14\farcs6, -78$^{\circ}$ & 20\farcs7, 14\farcs1, -78$^{\circ}$ & 18\farcs6, 12\farcs7, -79$^{\circ}$ \\
G209.29-19.65S1 & G209.29S1 & 19\farcs4, 9\farcs6, -73$^{\circ}$ & 21\farcs3, 14\farcs5, -80$^{\circ}$ & 21\farcs2, 14\farcs6, -78$^{\circ}$ & 20\farcs8, 14\farcs1, -78$^{\circ}$ & 18\farcs7, 12\farcs7, -79$^{\circ}$ \\
G209.29-19.65S2 & G209.29S2 & 19\farcs4, 9\farcs6, -73$^{\circ}$ & 21\farcs4, 14\farcs5, -80$^{\circ}$ & 21\farcs2, 14\farcs6, -78$^{\circ}$ & 20\farcs8, 14\farcs1, -78$^{\circ}$ & 18\farcs8, 12\farcs7, -79$^{\circ}$ \\
G209.55-19.68N2 & G209.55N2 & 19\farcs4, 9\farcs6, -73$^{\circ}$ & 21\farcs4, 14\farcs5, -80$^{\circ}$ & 21\farcs2, 14\farcs6, -78$^{\circ}$ & 20\farcs8, 14\farcs1, -78$^{\circ}$ & 18\farcs6, 12\farcs7, -79$^{\circ}$ \\
G209.77-19.40E3 & G209.77E3 & 19\farcs3, 9\farcs6, -73$^{\circ}$ & 21\farcs3, 14\farcs5, -80$^{\circ}$ & 21\farcs2, 14\farcs6, -78$^{\circ}$ & 20\farcs7, 14\farcs0, -78$^{\circ}$ & 18\farcs5, 12\farcs7, -79$^{\circ}$ \\
G209.94-19.52N  & G209.94N  & 19\farcs4, 9\farcs6, -73$^{\circ}$ & 21\farcs4, 14\farcs5, -80$^{\circ}$ & 21\farcs2, 14\farcs6, -79$^{\circ}$ & 20\farcs8, 14\farcs0, -78$^{\circ}$ & 18\farcs6, 12\farcs7, -80$^{\circ}$ \\
G210.37-19.53N  & G210.37N  & 19\farcs3, 9\farcs6, -73$^{\circ}$ & 21\farcs7, 11\farcs1, -76$^{\circ}$ & 21\farcs2, 14\farcs5, -79$^{\circ}$ & 20\farcs8, 14\farcs0, -79$^{\circ}$ & 18\farcs5, 12\farcs7, -80$^{\circ}$ \\
G210.82-19.47N2 & G210.82N2 & 19\farcs3, 9\farcs6, -74$^{\circ}$ & 21\farcs6, 11\farcs0, -74$^{\circ}$ & 21\farcs2, 14\farcs7, -80$^{\circ}$ & 20\farcs8, 14\farcs1, -80$^{\circ}$ & 18\farcs6, 12\farcs8, -81$^{\circ}$ \\
G211.16-19.33N4 & G211.16N4 & 19\farcs3, 9\farcs6, -74$^{\circ}$ & 21\farcs5, 14\farcs5, -82$^{\circ}$ & 21\farcs0, 10\farcs4, -74$^{\circ}$ & 20\farcs8, 14\farcs1, -80$^{\circ}$ & 18\farcs6, 9\farcs5, -76$^{\circ}$ \\
G211.16-19.33N5 & G211.16N5 & 19\farcs3, 9\farcs5, -74$^{\circ}$ & 21\farcs3, 13\farcs7, -79$^{\circ}$ & 21\farcs0, 10\farcs4, -74$^{\circ}$ & 20\farcs8, 14\farcs2, -80$^{\circ}$ & 19\farcs2, 9\farcs9, -78$^{\circ}$ \\
G212.10-19.15N1 & G212.10N1 & 19\farcs2, 9\farcs5, -74$^{\circ}$ & 23\farcs5, 14\farcs8, -79$^{\circ}$ & 22\farcs3, 14\farcs8, -77$^{\circ}$ & 22\farcs0, 10\farcs7, -72$^{\circ}$ & 20\farcs0, 9\farcs3, -72$^{\circ}$
\enddata
\tablecomments{MAJ, MIN, and PA are the major axis size, minor axis size, and position angle.}
\end{deluxetable}

\begin{deluxetable}{llcccccc}
\caption{\label{tab:appx:noise}Noise levels of the observations.}
\tablehead{
\colhead{Name} & \colhead{Short Name} & \colhead{3~mm} & \colhead{N$_2$H$^+$} & \colhead{CCS} & \colhead{CH$_3$OH (ACA)} & \colhead{c-C$_3$HD} & \colhead{CH$_3$OH (Yebes)} \\
\colhead{} & \colhead{} & \colhead{(mK)} & \colhead{(mK)} & \colhead{(mK)} & \colhead{(mK)} & \colhead{(mK)} & \colhead{(mK)} 
}
\startdata
G203.21-11.20E1 & G203.21E1 & 0.23 & 9.8 & 13.5 & 8.9 & 6.7 & 6.2 \\
G205.46-14.56M3 & G205.46M3 & 0.37 & 11.5 & 11.2 & 10.8 & 8.0 & 10.0 \\
G206.21-16.17N  & G206.21N & 0.26 & 11.2 & 12.0 & 11.0 & 8.1 & 15.6 \\
G206.21-16.17S  & G206.21S & 0.23 & 11.3 & 12.2 & 11.8 & 8.5 & 15.6 \\
G208.68-19.20N2 & G208.68N2 & 0.77 & 10.8 & 11.7 & 11.0 & 8.0 & 12.5 \\
G209.29-19.65N1 & G209.29N1 & 0.3 & 11.1 & 10.2 & 10.6 & 7.8 & 15.1 \\
G209.29-19.65S1 & G209.29S1 & 0.42 & 10.1 & 10.4 & 10.7 & 7.0 & 10.1 \\
G209.29-19.65S2 & G209.29S2 & 0.41 & 11.1 & 10.4 & 11.0 & 7.9 & 12.2 \\
G209.55-19.68N2 & G209.55N2 & 0.33 & 11.1 & 11.1 & 11.2 & 8.2 & 9.4 \\
G209.77-19.40E3 & G209.77E3 & 0.3 & 11.7 & 10.8 & 11.3 & 7.6 & 8.2 \\
G209.94-19.52N  & G209.94N & 0.33 & 11.2 & 10.5 & 10.5 & 7.2 & 9.6 \\
G210.37-19.53N  & G210.37N & 0.24 & 14.2 & 10.5 & 11.2 & 7.8 & 9.1 \\
G210.82-19.47N2 & G210.82N2 & 0.22 & 14.3 & 11.1 & 10.5 & 7.5 & 12.7 \\
G211.16-19.33N4 & G211.16N4 & 0.25 & 11.0 & 15.2 & 11.0 & 10.7 & 11.2 \\
G211.16-19.33N5 & G211.16N5 & 0.23 & 11.7 & 16.0 & 10.8 & 9.5 & 10.8 \\
G212.10-19.15N1 & G212.10N1 & 0.3 & 10.0 & 11.6 & 12.8 & 9.9 & 13.2
\enddata
\end{deluxetable}

\resetapptablenumbers
\section{Source Overview \label{appx:overview}}

\subsection{Literature Review \label{appx:overview:lit}}
All targets in this study were originally reported in \citet{2018Yi_PGCC_SCUBA2_II}, which compared the overall properties of Planck Galactic Cold Clumps (PGCCs) in Orion A, Orion B, and the $\lambda$ Orionis cloud based on 850 \micron\ dust continuum data from SCUBA-2 on the JCMT.
As part of the sample, most of the cores were later observed by \citet{2020Kim_PGCC_45m} at 76--94 GHz using the Nobeyama 45~m telescope for molecules such as N$_2$D$^+$, N$_2$H$^+$, DNC, HN$^{13}$C, CCS, HC$_3$N, and c-C$_3$H$_2$. 
Subsequently, these clumps were observed by ALMA as part of the ALMASOP project \citep{2020Dutta_ALMASOP}, which further identified substructures within each clump.
Following their inclusion in the ALMASOP catalog, (part of) this sample has been used in several studies, which are described as follows and summarized in Table ~\ref{tab:lit}. 
\citet{2021Sahu_ALMASOP_presstellar} presented a detailed analysis of the five densest starless cores, which show centrally concentrated regions with sizes of $\sim$ 2,000~au and average densities of a few $\times$10$^7$~cm$^{-3}$.
\citet{2021Yi_PGCC_chem} observed 80 cores (including nine starless cores in this study), distributed across Orion A, Orion B, and the $\lambda$ Orionis clouds, and investigated the relationship between ultraviolet (UV) radiation fields and chemical composition in young stellar objects (YSOs).
\citet{2021Tatematsu_SCOPE} observed 107 starless and prestellar cores, including six from this study, across Orion A, Orion B, and the $\lambda$ Orionis clouds, investigating the emission lines of N$_2$H$^+$, HC$_3$N, and CCS with the Nobeyama 45-m telescope.
\citet{2022Tatematsu_ALMASOP_inward} studied inward motions in 30 starless (including all 16 cores in this study) and six protostellar cores using the data obtained from the Nobeyama 45~m radio telescope observations.
The five sources in \citet{2023Sahu_ALMASOP_density} and \citet{2025Hsu_ALMASOP_starless} were adopted from \citet{2021Sahu_ALMASOP_presstellar}, representing the densest starless cores in the ALMASOP sample.
\citet{2023Sahu_ALMASOP_density} modeled the density structures using observations from ALMA, ACA, and JCMT SCUBA-2.
\citet{2025Hsu_ALMASOP_starless} analyzed newly retrieved ALMA Band 3 data, discovered a shock interface between N$_2$H$^+$ and CH$_3$OH, and proposed a turbulence-induced mass assembly framework, supporting the idea that turbulence drives mass accumulation and causes fragmented CH$_3$OH morphologies in starless cores.
We note that the source G205.46–14.56M3 is referred to as G205.46–14.56North1 in \citet{2018Yi_PGCC_SCUBA2_II}.
Additional references focused on individual sources are discussed in the contextual descriptions of each object.

\begin{deluxetable}{llccccccccc}
\caption{\label{tab:lit} Summary of sources included in literature. The ``\checkmark'' labels the the sources included in the corresponding literature}
\tablehead{\colhead{Name} & \colhead{Short Name} & \colhead{Yi+18} & \colhead{Kim+20} & \colhead{Dutta+20} & \colhead{Sahu+21} & \colhead{Yi+21} & \colhead{Tatematsu+21} & \colhead{Tatematsu+22} & \colhead{Sahu+23} & \colhead{Hsu+25}}
\startdata
G203.21-11.20E1 & G203.21E1 & \checkmark & \checkmark  & \checkmark &   & \checkmark & \checkmark & \checkmark &   & \\
G205.46-14.56M3 & G205.46M3 & \checkmark & \checkmark  & \checkmark & \checkmark &   &   & \checkmark & \checkmark & \checkmark \\
G206.21-16.17N  & G206.21N & \checkmark & \checkmark  & \checkmark &   & \checkmark &   & \checkmark &   & \\
G206.21-16.17S  & G206.21S & \checkmark &            & \checkmark &   & \checkmark &   & \checkmark &   & \\
G208.68-19.20N2 & G208.68N2 & \checkmark & \checkmark  & \checkmark & \checkmark & \checkmark & \checkmark & \checkmark & \checkmark & \checkmark \\
G209.29-19.65N1 & G209.29N1 & \checkmark & \checkmark  & \checkmark &   &   &   & \checkmark &   & \\
G209.29-19.65S1 & G209.29S1 & \checkmark & \checkmark  & \checkmark & \checkmark &   &   & \checkmark & \checkmark & \checkmark \\
G209.29-19.65S2 & G209.29S2 & \checkmark & \checkmark  & \checkmark &   &   & \checkmark & \checkmark &   & \\
G209.55-19.68N2 & G209.55N2 & \checkmark & \checkmark  & \checkmark &   & \checkmark &   & \checkmark &   & \\
G209.77-19.40E3 & G209.77E3 & \checkmark & \checkmark  & \checkmark &   & \checkmark &   & \checkmark &   & \\
G209.94-19.52N  & G209.94N & \checkmark & \checkmark  & \checkmark & \checkmark & \checkmark & \checkmark & \checkmark & \checkmark & \checkmark \\
G210.37-19.53N  & G210.37N & \checkmark & \checkmark  & \checkmark &   &   &   & \checkmark &   & \\
G210.82-19.47N2 & G210.82N2 & \checkmark & \checkmark  & \checkmark &   &   &   & \checkmark &   & \\
G211.16-19.33N4 & G211.16N4 & \checkmark & \checkmark  & \checkmark &   & \checkmark &   & \checkmark &   & \\
G211.16-19.33N5 & G211.16N5 & \checkmark & \checkmark  & \checkmark &   & \checkmark & \checkmark & \checkmark &   & \\
G212.10-19.15N1 & G212.10N1 & \checkmark & \checkmark  & \checkmark & \checkmark &   & \checkmark & \checkmark & \checkmark & \checkmark \\
\enddata
\tablerefs{Yi+18: \citet{2018Yi_PGCC_SCUBA2_II}; Kim+20: \citet{2020Kim_PGCC_45m}; Dutta+20: \citet{2020Dutta_ALMASOP}; Sahu+21: \citet{2021Sahu_ALMASOP_presstellar}; Yi+21: \citet{2021Yi_PGCC_chem}; Tatematsu+21: \citet{2021Tatematsu_SCOPE};  Tatematsu+22: \citet{2022Tatematsu_ALMASOP_inward}; Sahu+23: \citet{2023Sahu_ALMASOP_density}; Hsu+25: \citet{2025Hsu_ALMASOP_starless}; }
\end{deluxetable}

\subsection{Integrated Intensity Images and Column Densities}
\label{appx:overview:mom0}

\begin{figure}[h]\centering\includegraphics[width=.99\textwidth]{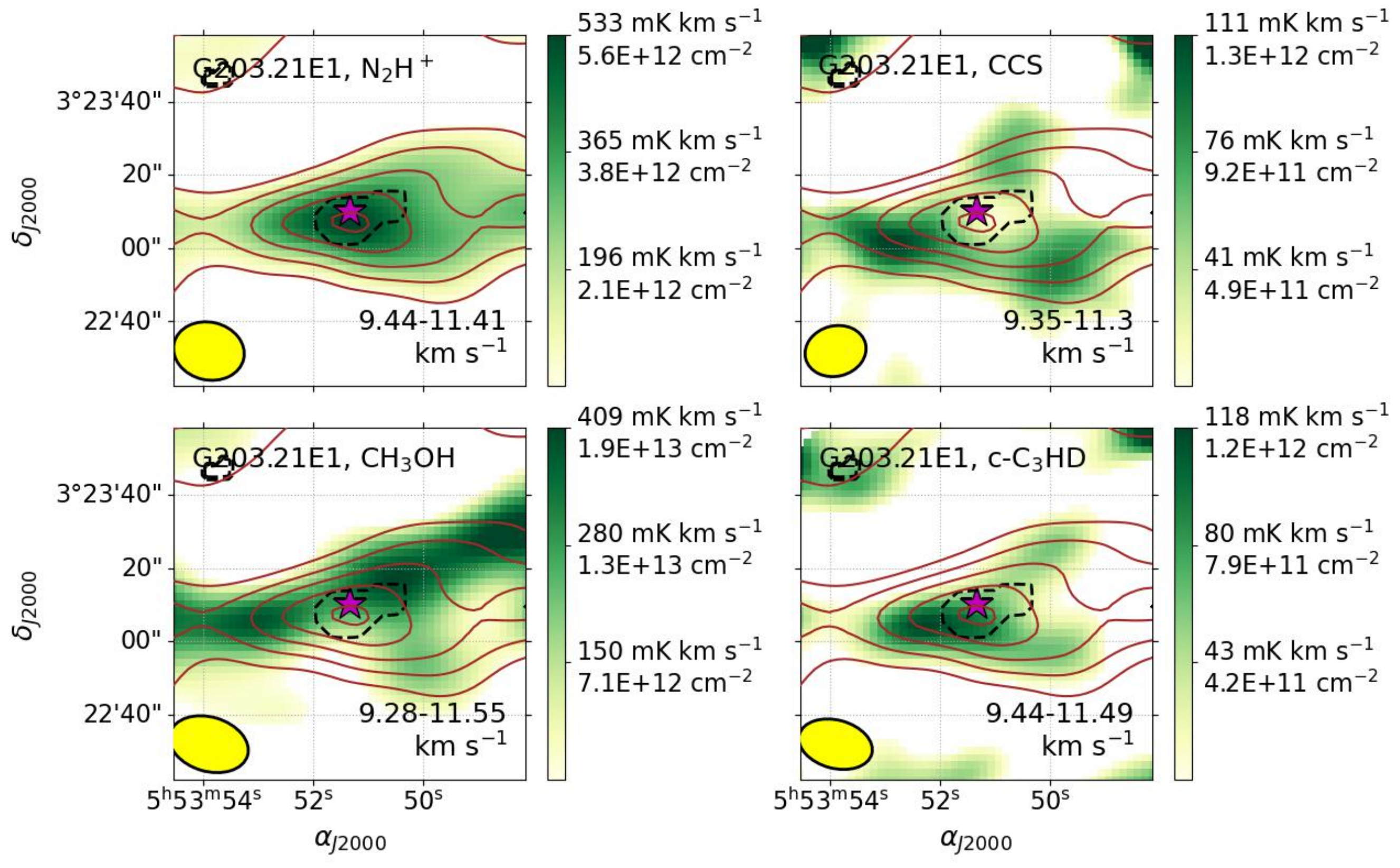}\caption{\label{fig:appx:mom0_G203.21E1}Integrated intensity images of G203.21-11.20E1 (G203.21E1). Please see the context in Appendix \ref{appx:overview:mom0} for the captions.}\end{figure}

\begin{figure}[h]\centering\includegraphics[width=.99\textwidth]{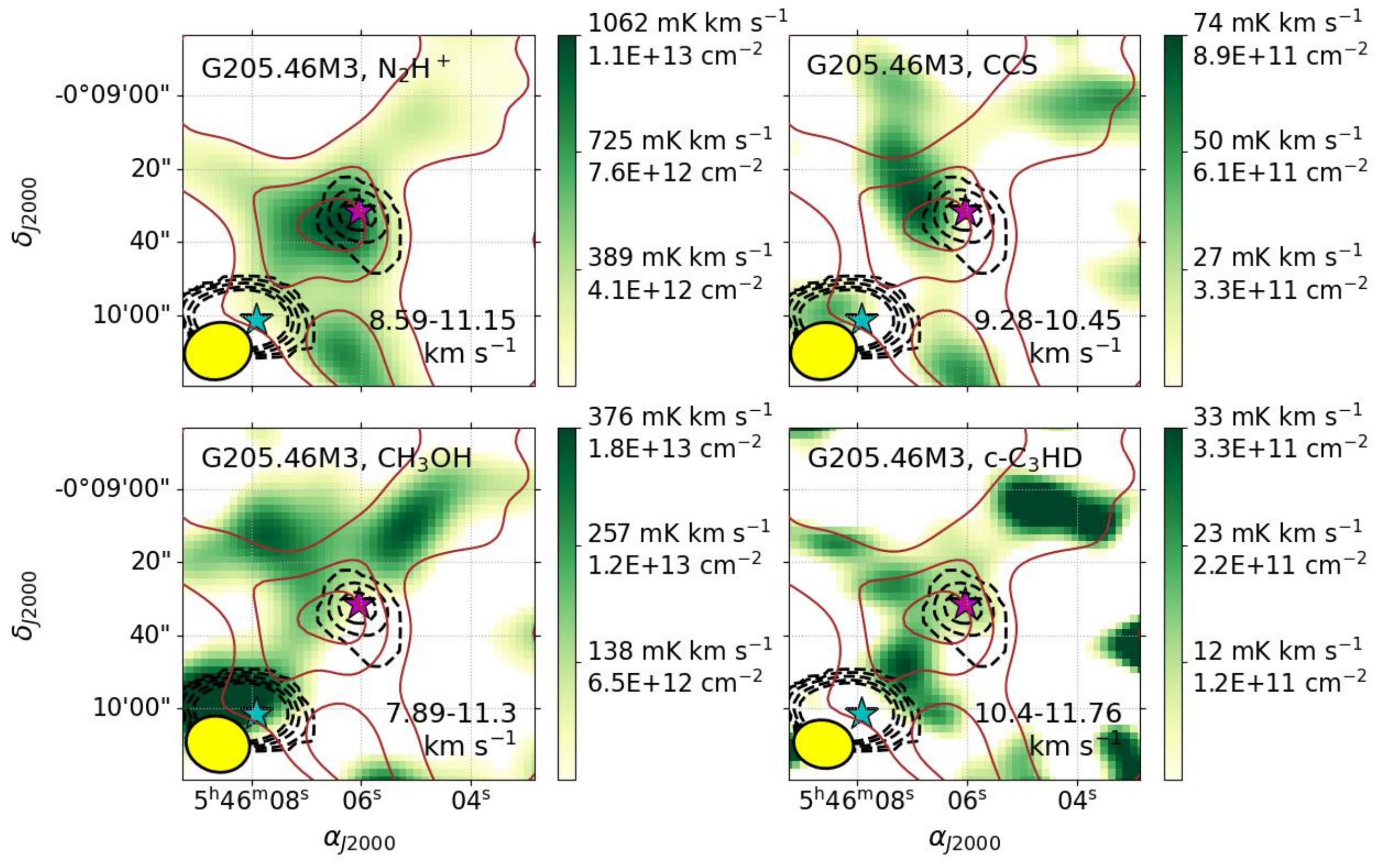}\caption{\label{fig:appx:mom0_G205.46M3}Integrated intensity images of G205.46-14.56M3 (G205.46M3). The texts on each panel green rasters label the source short name and the molecular species of the images. Please see the context in Appendix \ref{appx:overview:mom0} for the captions.}\end{figure}

\begin{figure}[h]\centering\includegraphics[width=.99\textwidth]{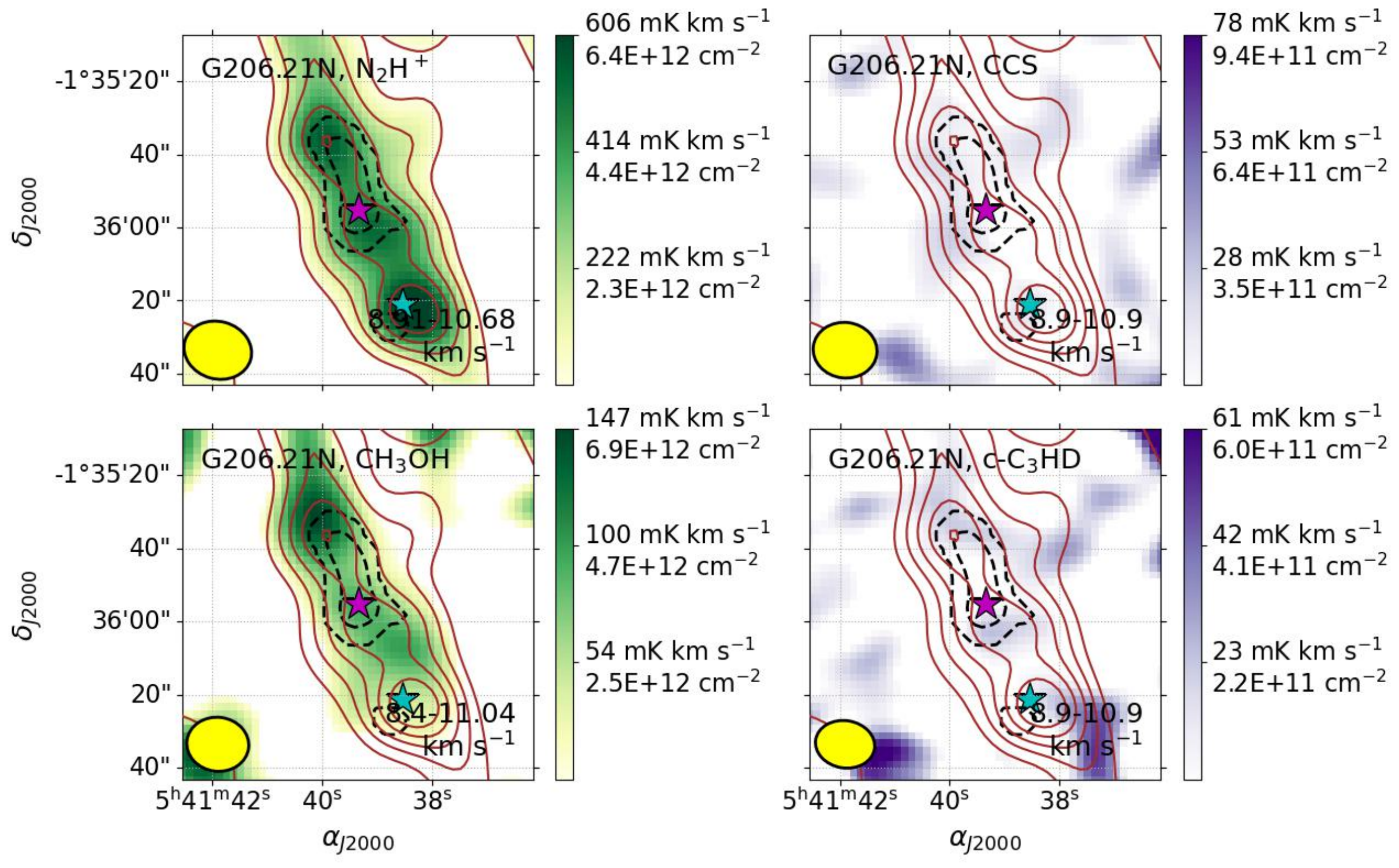}\caption{\label{fig:appx:mom0_G206.21N}Integrated intensity images of G206.21-16.17N (G206.21N). Please see the context in Appendix \ref{appx:overview:mom0} for the captions.}\end{figure}

\begin{figure}[h]\centering\includegraphics[width=.99\textwidth]{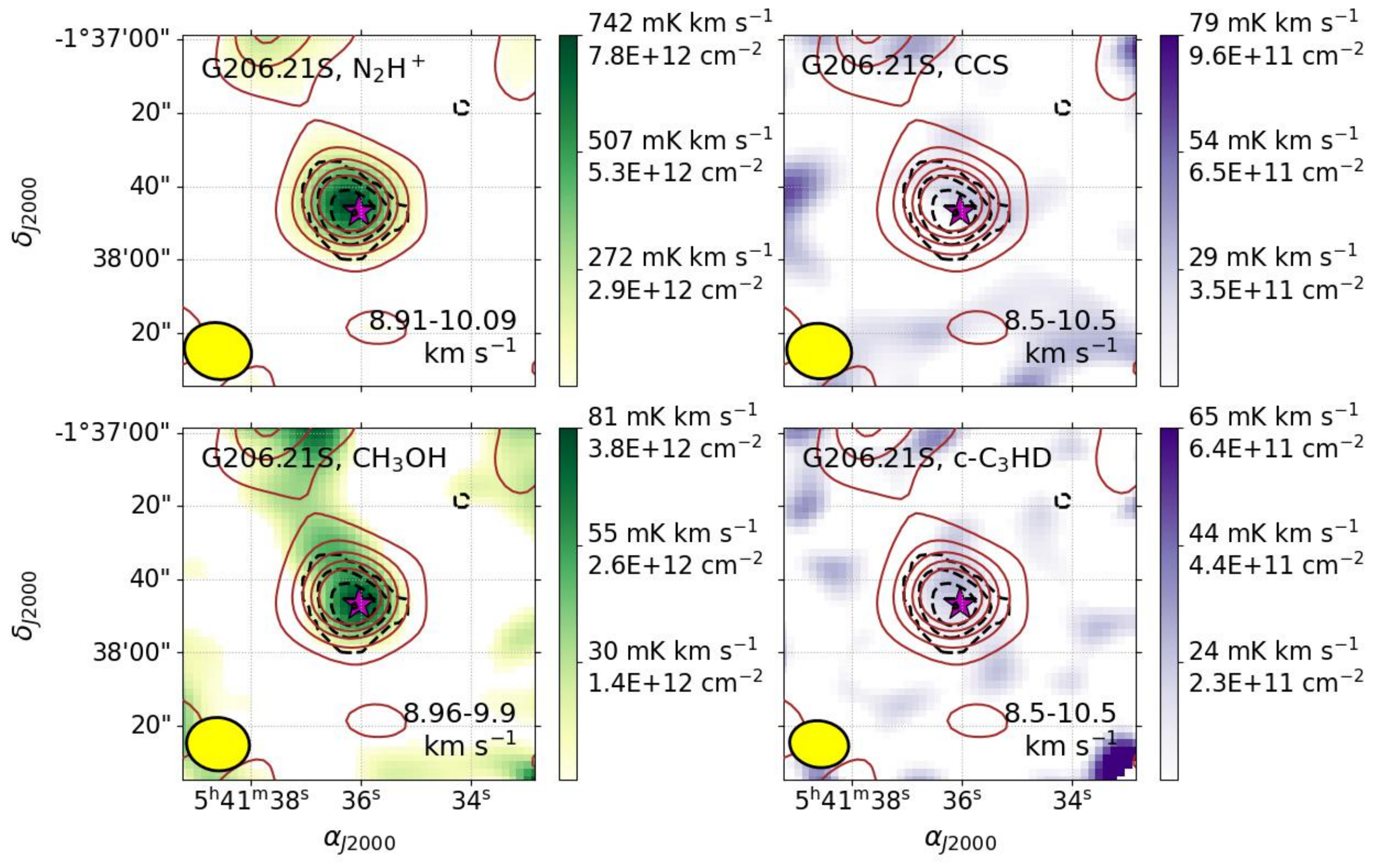}\caption{\label{fig:appx:mom0_G206.21S}Integrated intensity images of G206.21-16.17S (G206.21S). Please see the context in Appendix \ref{appx:overview:mom0} for the captions.}\end{figure}

\begin{figure}[h]\centering\includegraphics[width=.99\textwidth]{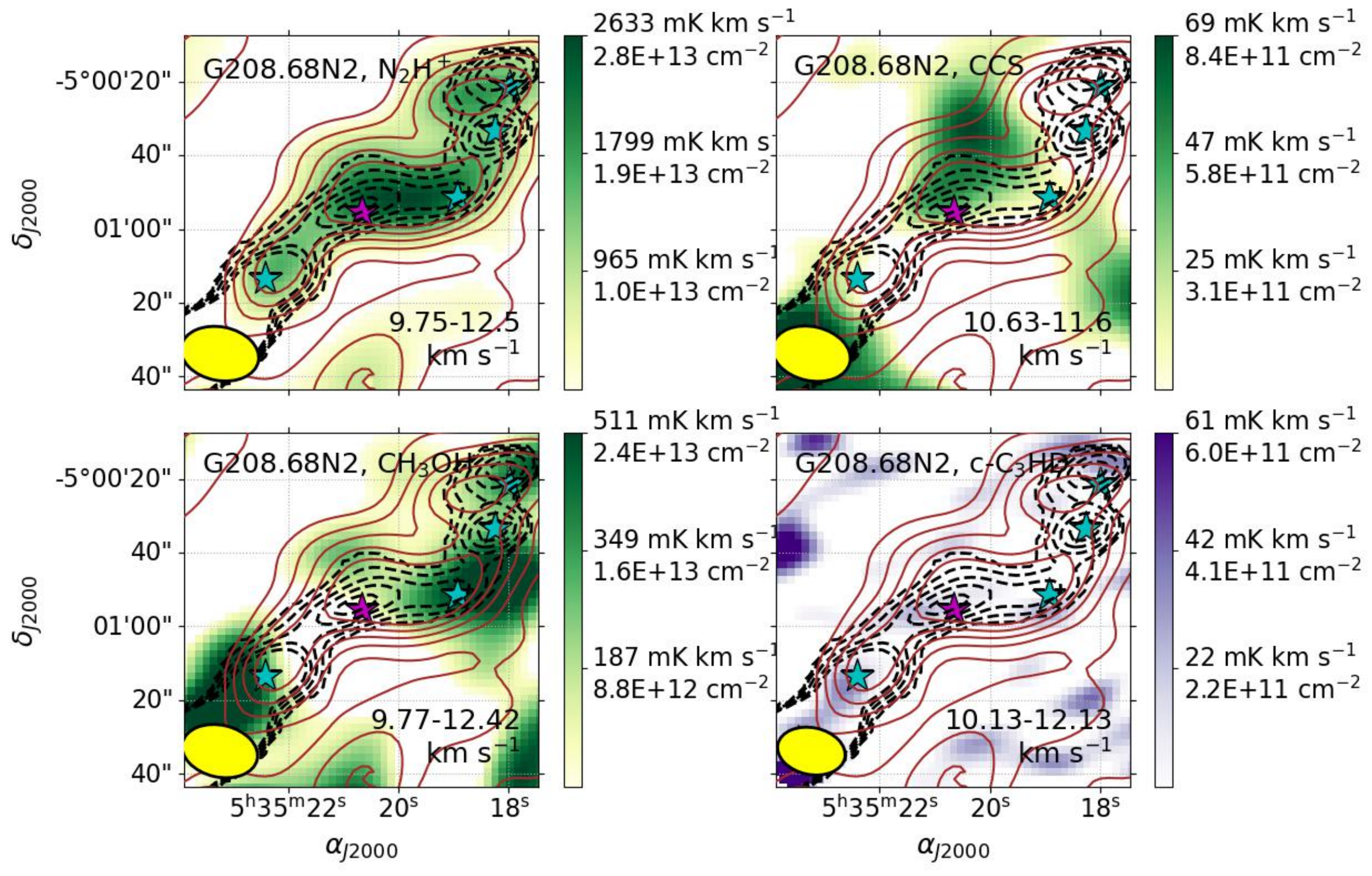}\caption{\label{fig:appx:mom0_G208.68N2}Integrated intensity images of G208.68-19.20N2 (G208.68N2). Please see the context in Appendix \ref{appx:overview:mom0} for the captions.}\end{figure}

\begin{figure}[h]\centering\includegraphics[width=.99\textwidth]{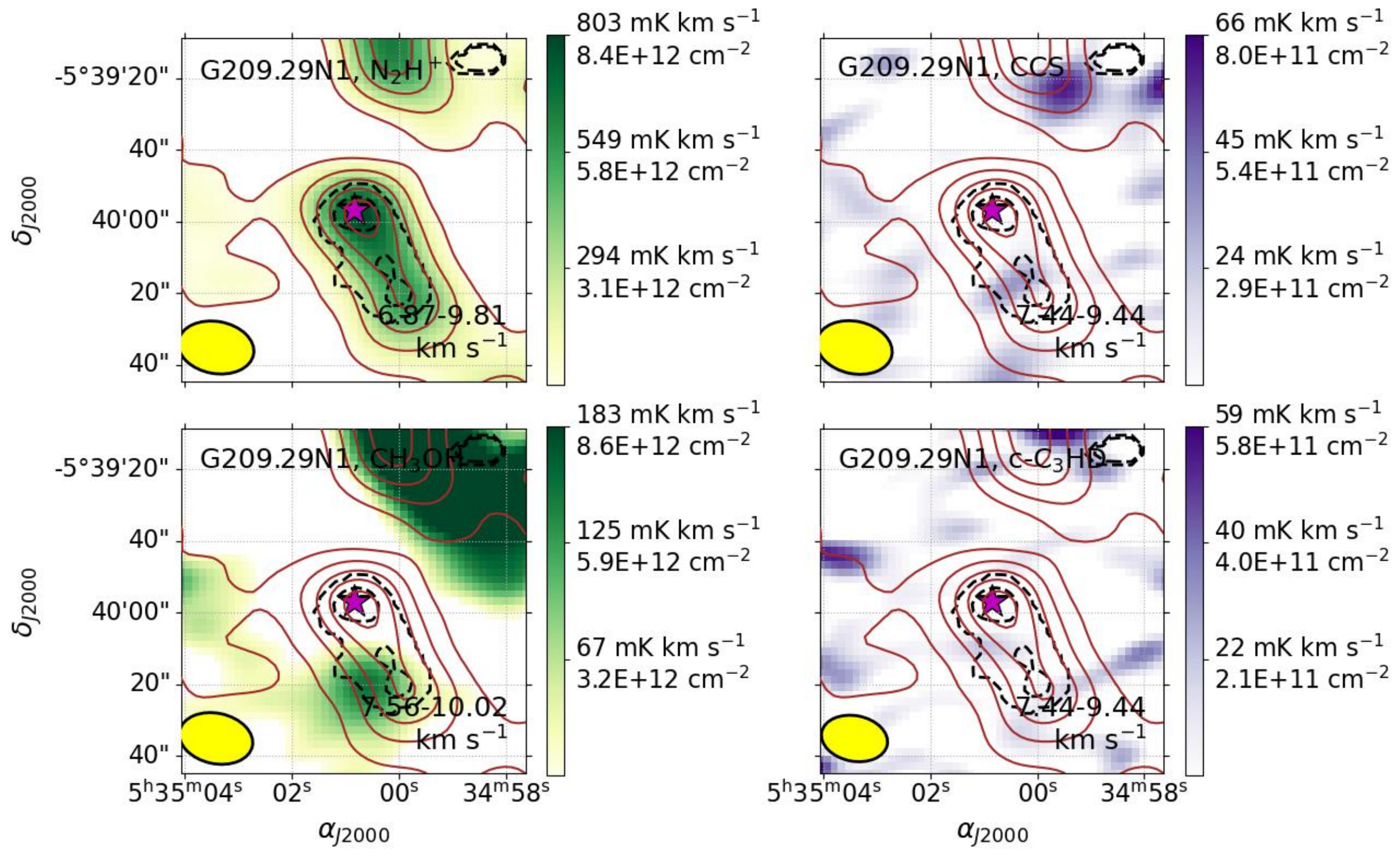}\caption{\label{fig:appx:mom0_G209.29N1}Integrated intensity images of G209.29-19.65N1 (G209.29N1). Please see the context in Appendix \ref{appx:overview:mom0} for the captions.}\end{figure}

\begin{figure}[h]\centering\includegraphics[width=.99\textwidth]{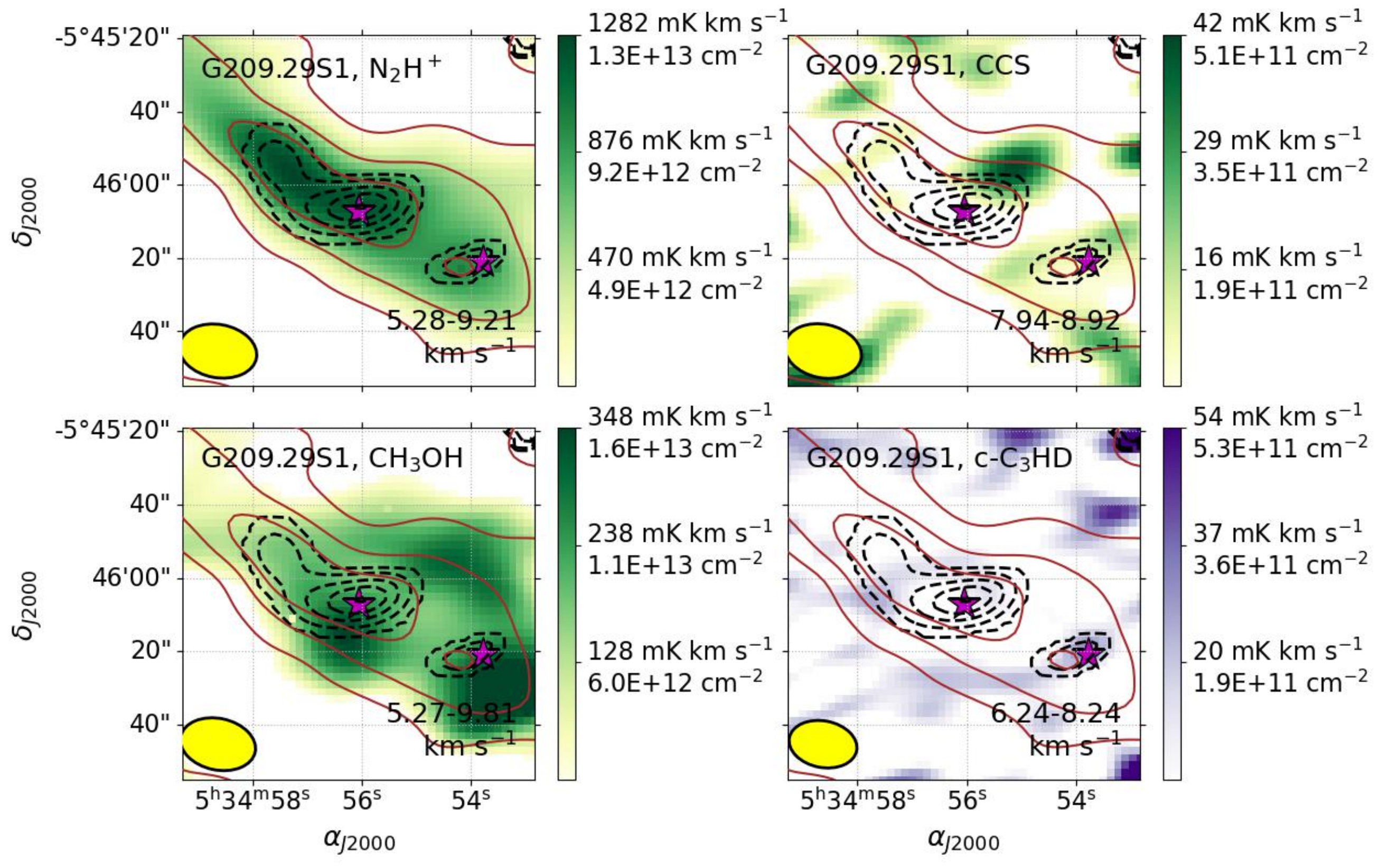}\caption{\label{fig:appx:mom0_G209.29S1}Integrated intensity images of G209.29-19.65S1 (G209.29S1). Please see the context in Appendix \ref{appx:overview:mom0} for the captions.}\end{figure}

\begin{figure}[h]\centering\includegraphics[width=.99\textwidth]{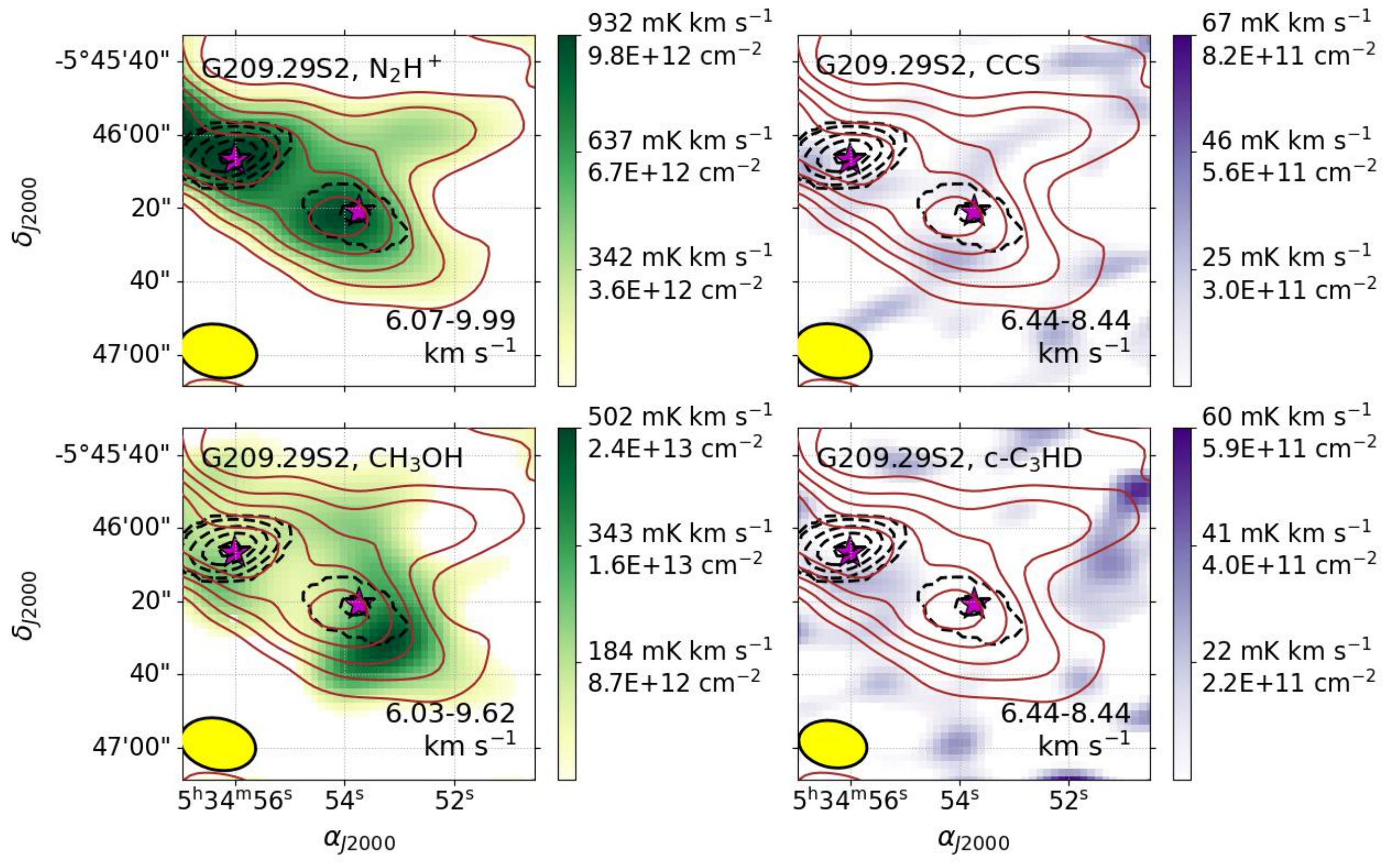}\caption{\label{fig:appx:mom0_G209.29S2}Integrated intensity images of G209.29-19.65S2 (G209.29S2). Please see the context in Appendix \ref{appx:overview:mom0} for the captions.}\end{figure}

\begin{figure}[h]\centering\includegraphics[width=.99\textwidth]{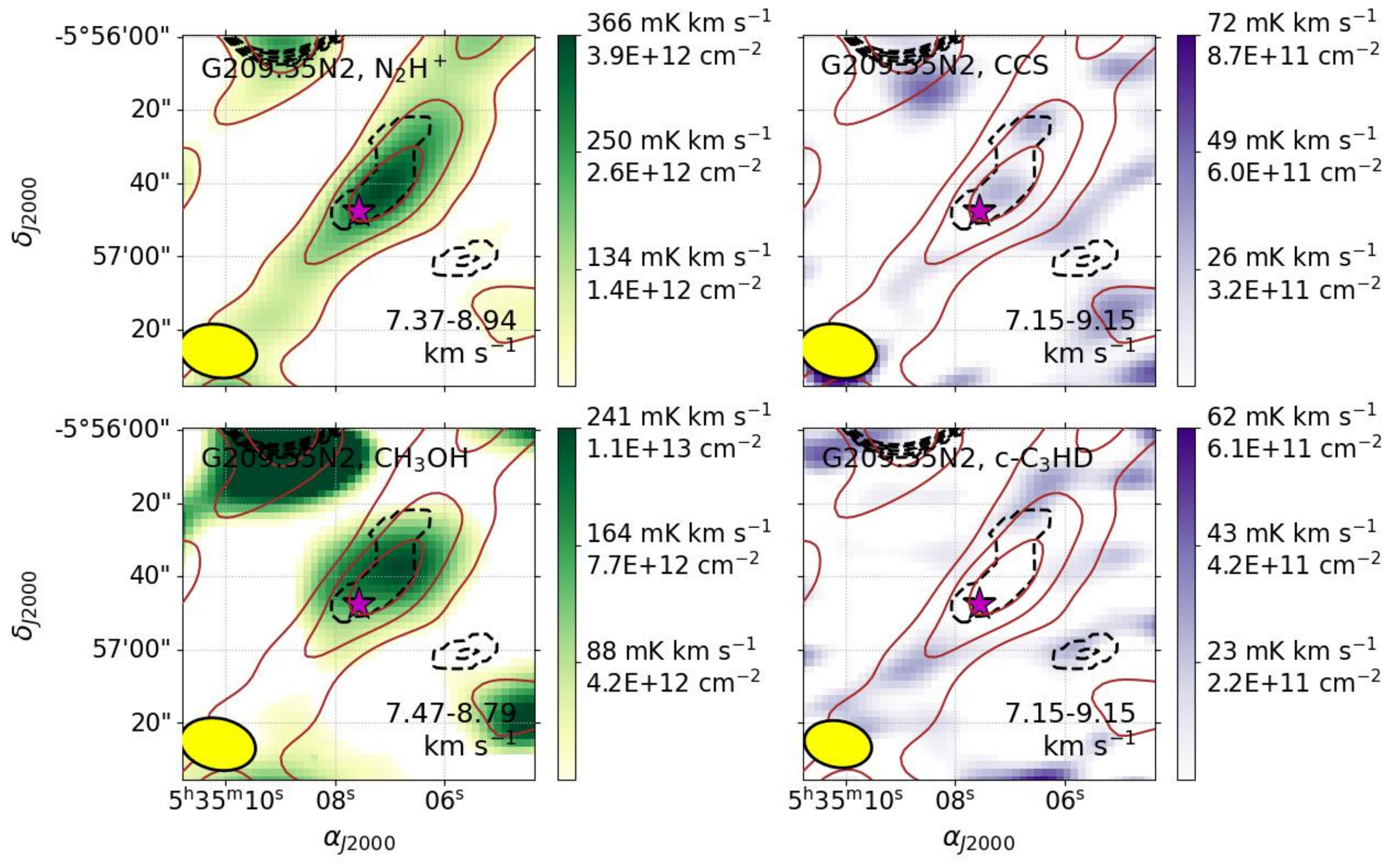}\caption{\label{fig:appx:mom0_G209.55N2}Integrated intensity images of G209.55-19.68N2 (G209.55N2). Please see the context in Appendix \ref{appx:overview:mom0} for the captions.}\end{figure}

\begin{figure}[h]\centering\includegraphics[width=.99\textwidth]{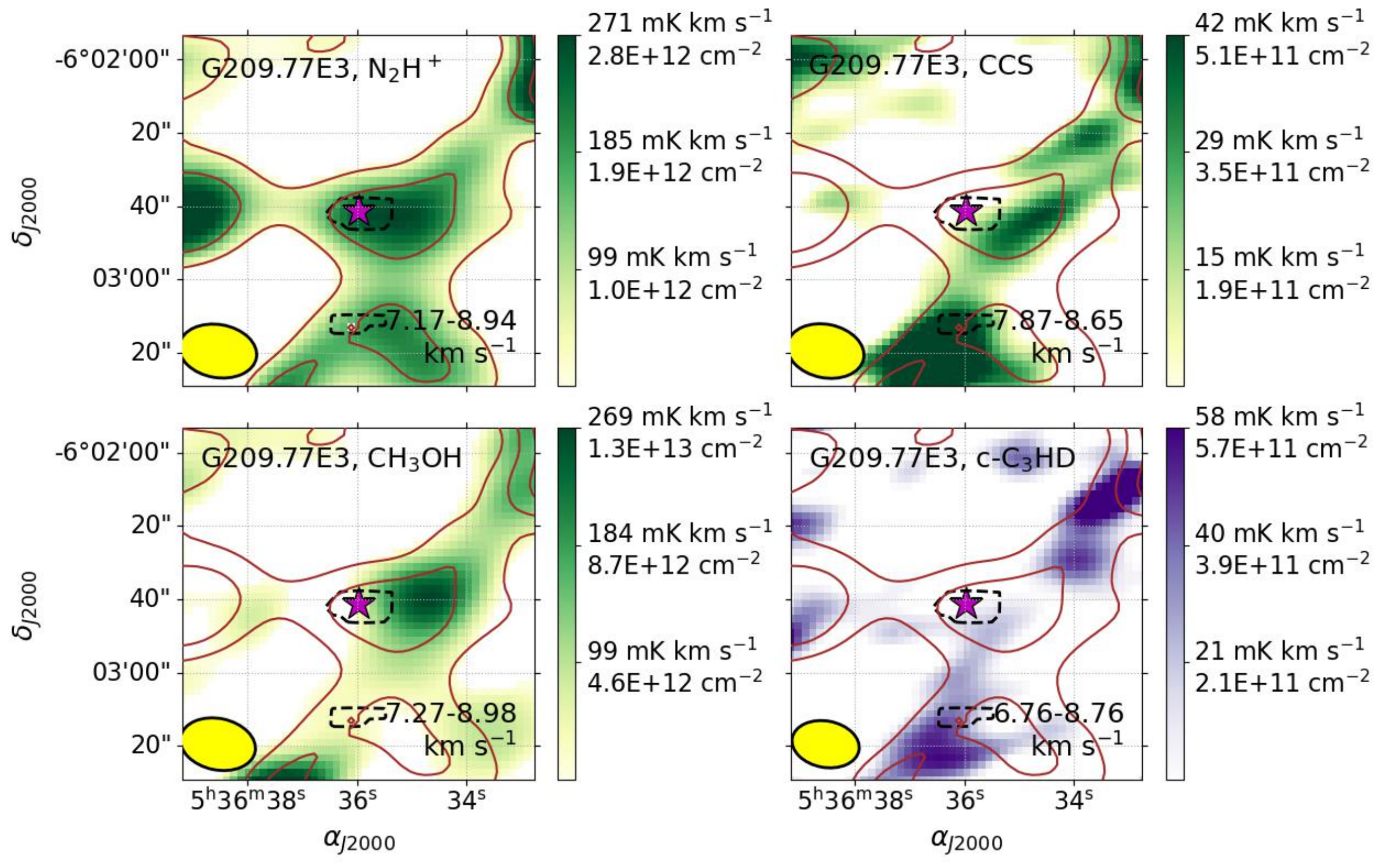}\caption{\label{fig:appx:mom0_G209.77E3}Integrated intensity images of G209.77-19.40E3 (G209.77E3). Please see the context in Appendix \ref{appx:overview:mom0} for the captions.}\end{figure}

\begin{figure}[h]\centering\includegraphics[width=.99\textwidth]{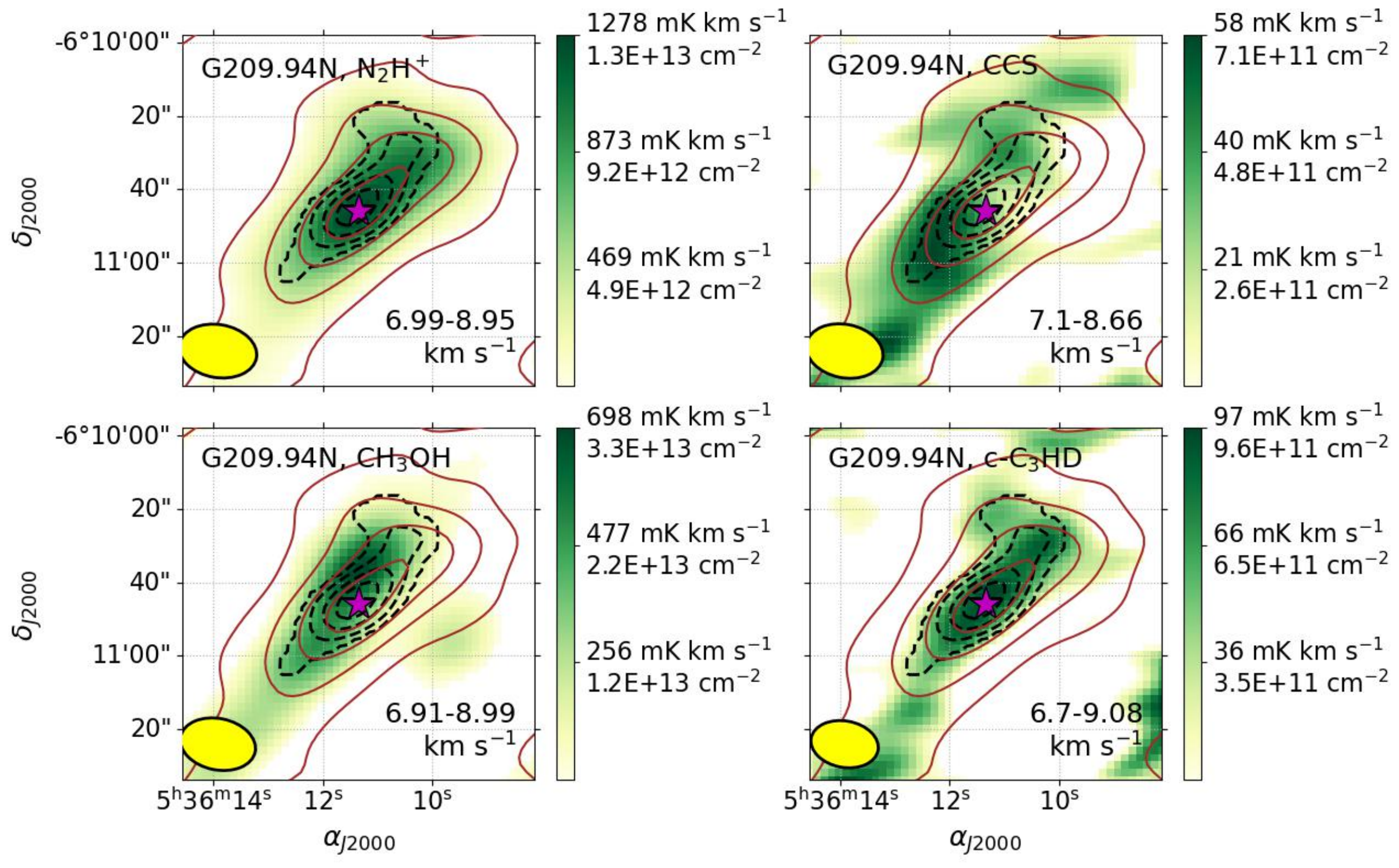}\caption{\label{fig:appx:mom0_G209.94N}Integrated intensity images of G209.94-19.52N (G209.94N). Please see the context in Appendix \ref{appx:overview:mom0} for the captions.}\end{figure}

\begin{figure}[h]\centering\includegraphics[width=.99\textwidth]{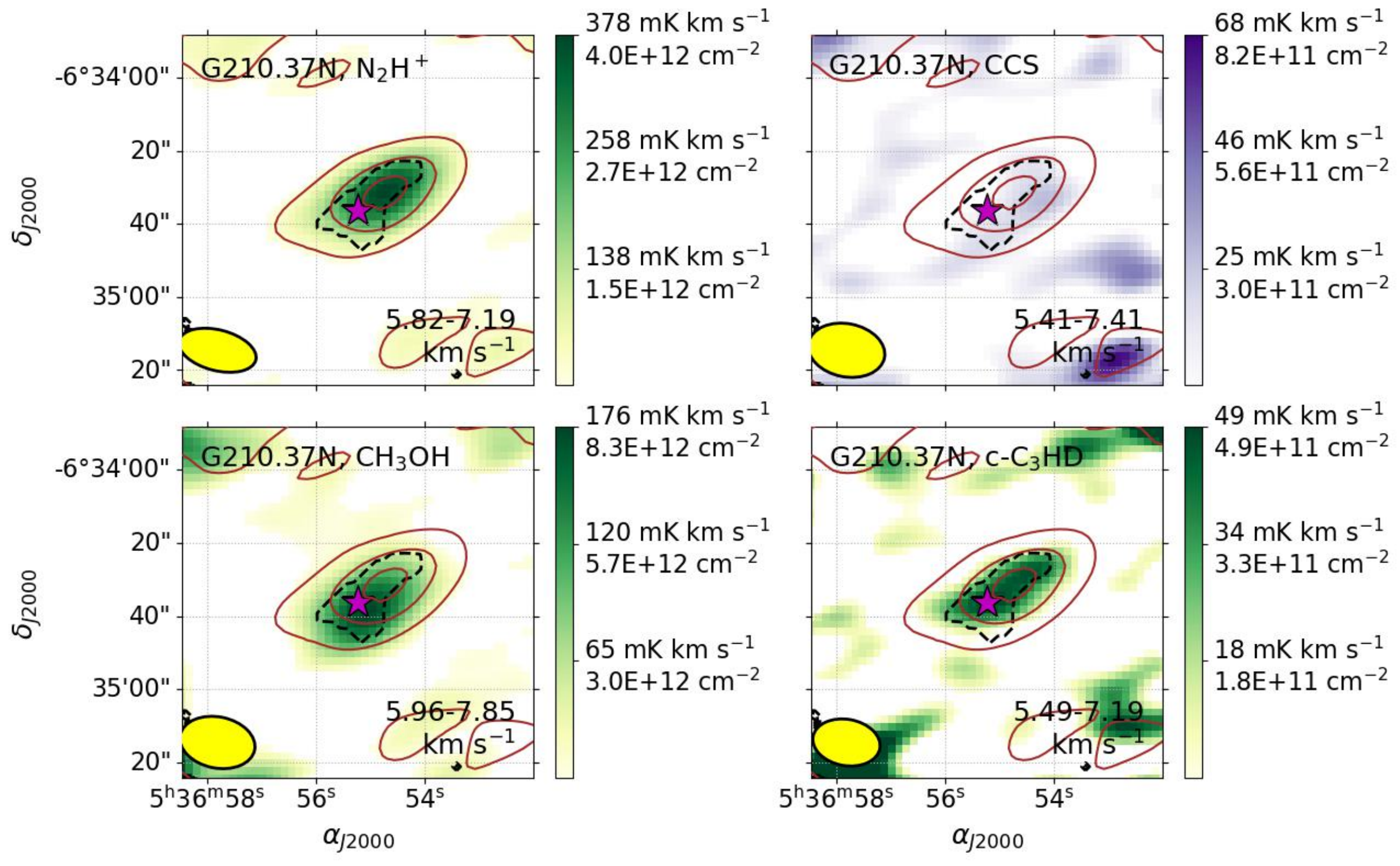}\caption{\label{fig:appx:mom0_G210.37N}Integrated intensity images of G210.37-19.53N (G210.37N). Please see the context in Appendix \ref{appx:overview:mom0} for the captions.}\end{figure}

\begin{figure}[h]\centering\includegraphics[width=.99\textwidth]{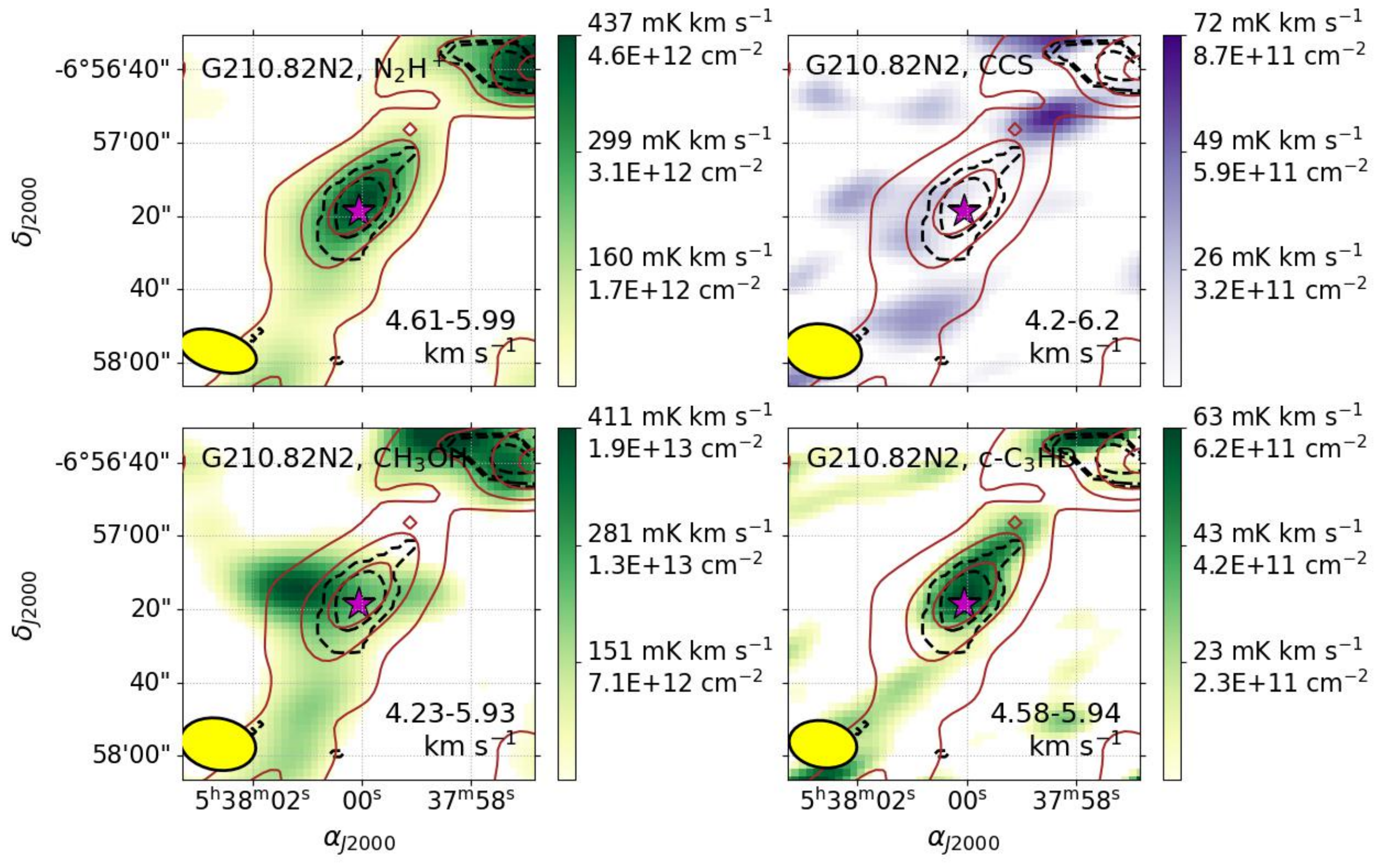}\caption{\label{fig:appx:mom0_G210.82N2}Integrated intensity images of G210.82-19.47N2 (G210.82N2). Please see the context in Appendix \ref{appx:overview:mom0} for the captions.}\end{figure}

\begin{figure}[h]\centering\includegraphics[width=.99\textwidth]{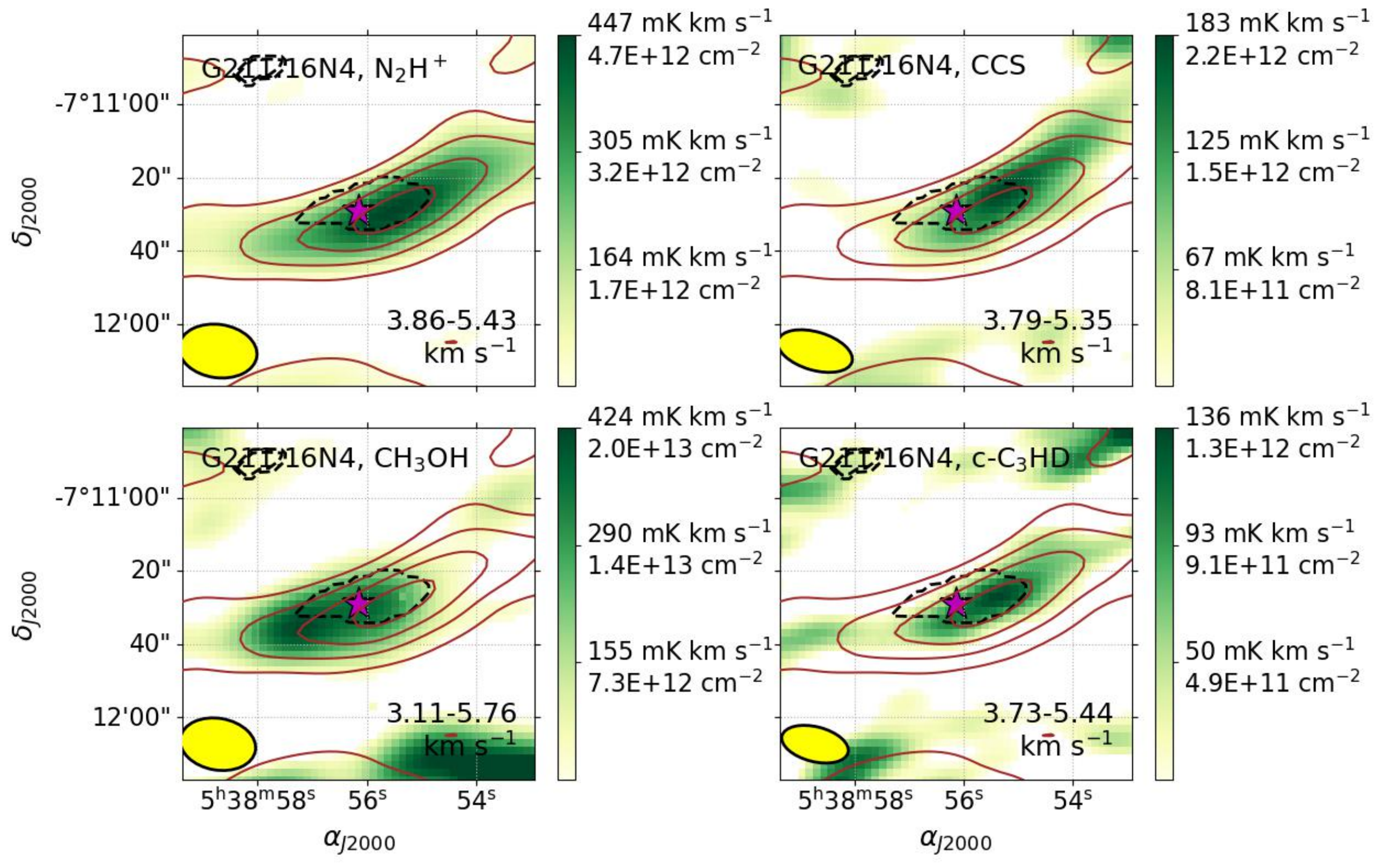}\caption{\label{fig:appx:mom0_G211.16N4}Integrated intensity images of G211.16-19.33N4 (G211.16N4). Please see the context in Appendix \ref{appx:overview:mom0} for the captions.}\end{figure}

\begin{figure}[h]\centering\includegraphics[width=.99\textwidth]{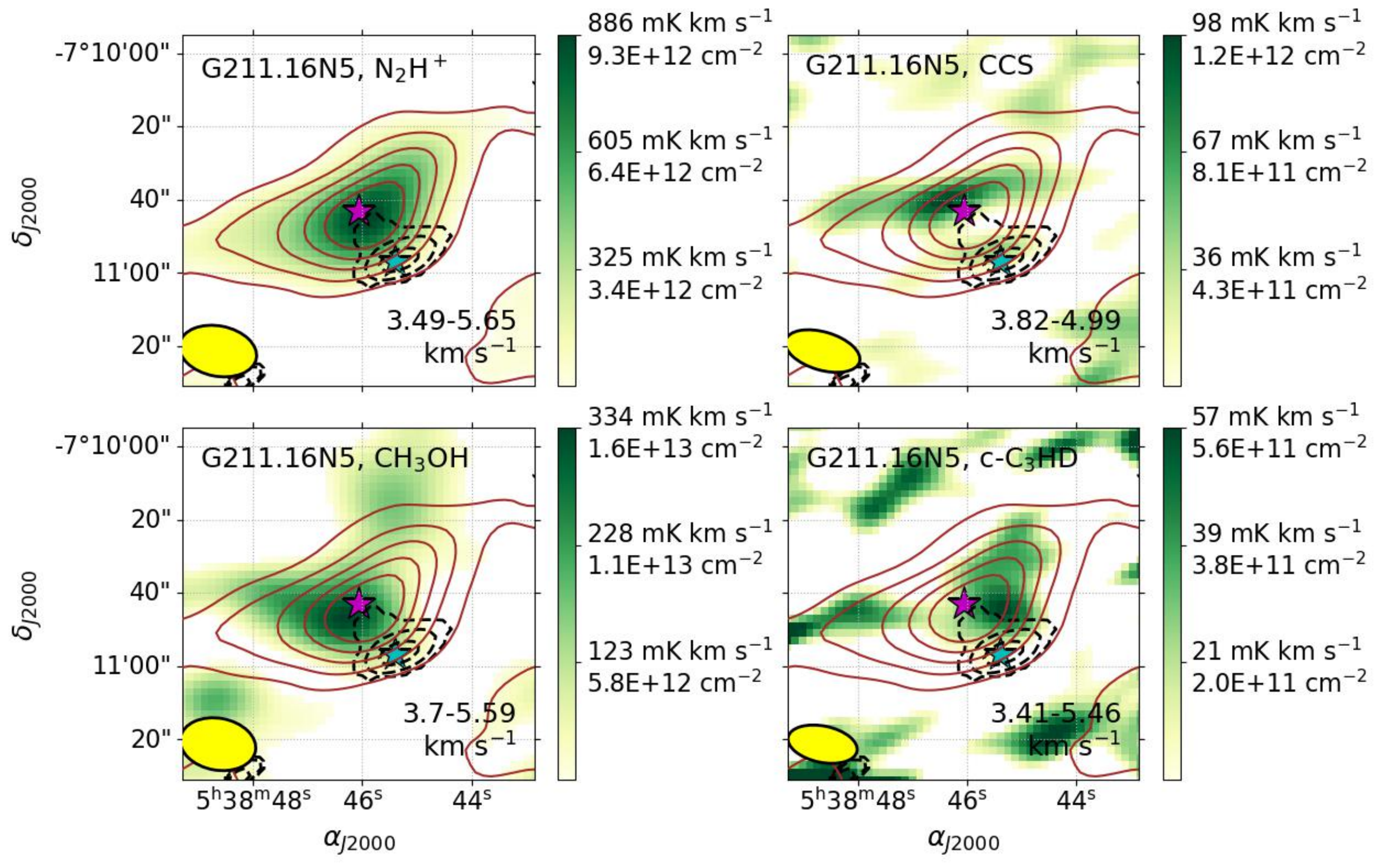}\caption{\label{fig:appx:mom0_G211.16N5}Integrated intensity images of G211.16-19.33N5 (G211.16N5). Please see the context in Appendix \ref{appx:overview:mom0} for the captions.}\end{figure}

\begin{figure}[h]\centering\includegraphics[width=.99\textwidth]{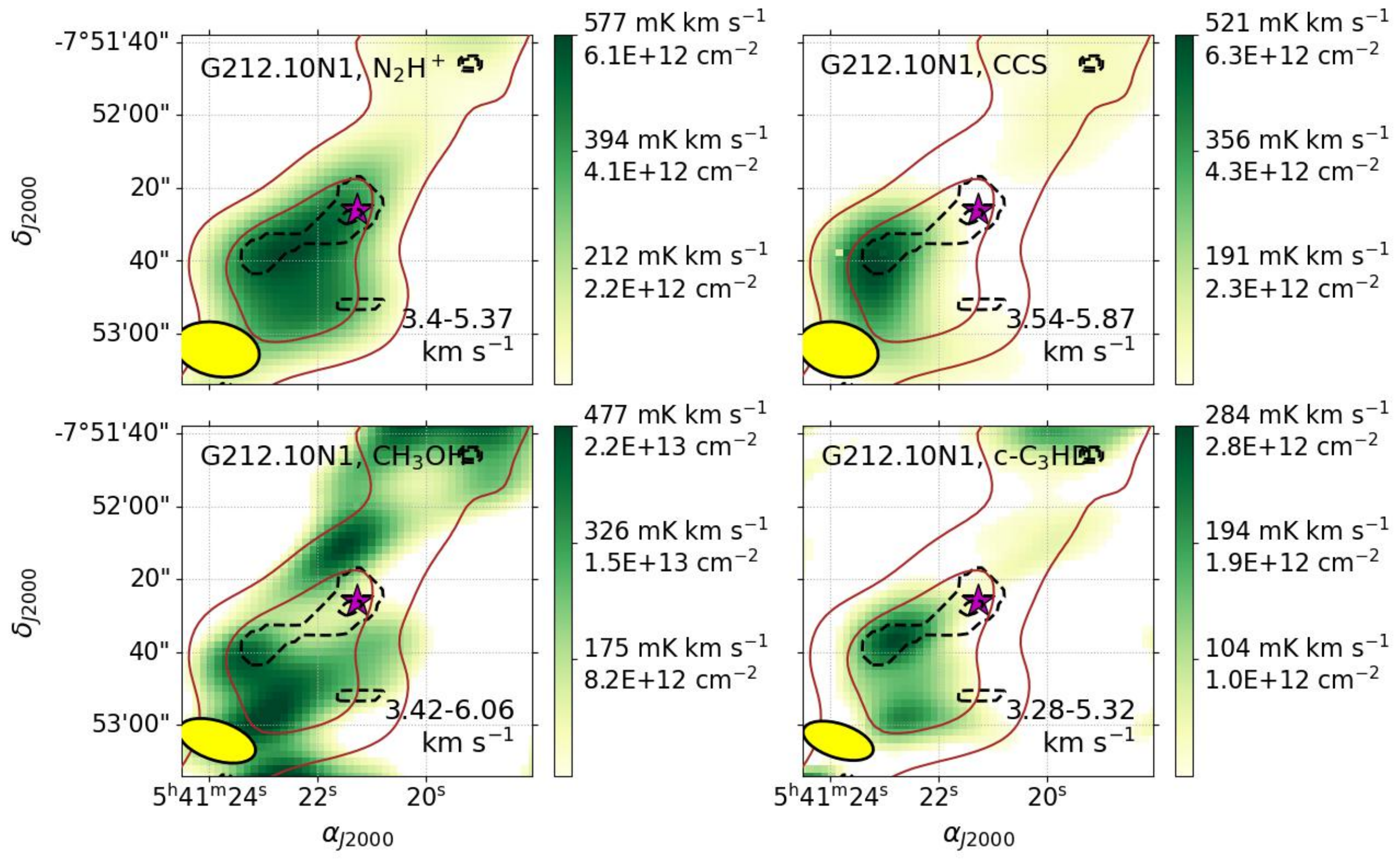}\caption{\label{fig:appx:mom0_G212.10N1}Integrated intensity images of G212.10-19.15N1 (G212.10N1). Please see the context in Appendix \ref{appx:overview:mom0} for the captions.}\end{figure}

Figures~\ref{fig:appx:mom0_G203.21E1}--\ref{fig:appx:mom0_G212.10N1} show the integrated intensity maps of N$_2$H$^+$, CH$_3$OH, CCS, and c-C$_3$HD for each source. 
Table~\ref{tab:appx:Ntot} lists the velocity integration interval we used. 
Rasters are colored green for detected molecules and purple for non-detections.
Brown contours and black dashed contours represent the N$_2$H$^+$ integrated intensity and the 3~mm dust continuum, respectively.
For five particularly dense starless cores, G205.46M3, G208.68N2, G209.94N, G209.29S1, and G212.10N1, the N$_2$H$^+$ contour levels are set at [5, 55, 105, 155, 205]$\sigma$.
For the remaining 11 cores, the contour levels are [5, 25, 45, 65, 85]$\sigma$.
The dust continuum contours are plotted at levels of [3, 6, 9, 12, 15]$\sigma$.
The cyan markers indicate known nearby sources reported by literature.
The magenta markers indicate the sources included in this study and the coordinates are listed in Table~\ref{tab:coord}. 
Each panel includes two color bar labels: the upper label indicates the integrated intensity, while the lower label shows the corresponding column density.

The column densities were calculated as follows. 
First, assuming optically thin and local thermodynamic equilibrium (LTE), we derive the upper upper state column density $N_\mathrm{u}$ from the integrated intensity of transition in terms of the products of brightness temperature and velocity $W$ via:
\begin{equation}
    N_\mathrm{u} = \frac{8 \pi k_\mathrm{B} \nu^2}{h c^3 A_\mathrm{ul}} W, 
\end{equation}
where $\nu$ is the frequency and $A_\mathrm{ul}$ is the Einstein A coefficient of the transition. 
We then derived the total column density $N_\mathrm{tot}$ via:
\begin{equation}
    \ln\left ( \frac{N_\mathrm{u}}{g_\mathrm{u}} \right ) =  \ln\left ( \frac{N_\mathrm{tot}}{Z} \right ) - \frac{E_\mathrm{u}}{k_\mathrm{B} T_\mathrm{rot}}, 
\end{equation}
where $g_\mathrm{u}$ is the upper state degeneracy, $T_\mathrm{rot}$ is the rotational temperature, and $Z$ is the partition function at $T_\mathrm{rot}$. 
Table~\ref{tab:appx:Ntot} lists the peak of the derived column density ($N^{p}_\mathrm{tot}$) within a 25\arcsec\ (10,000~au) circular aperture centered on the target position.

\begin{deluxetable}{ll|cc|cc|cc|cc}
\tablecaption{\label{tab:appx:Ntot}
The velocity integration interval ($\Delta v$) used to derive the integrated intensity maps and the peak of the resulting column density ($N^{p}_\mathrm{tot}$) within 25\arcsec\ (10,000~au).
The integration interval ($\Delta v$) is defined as the longest continuous range of velocity channels where the signal exceeds three times the rms noise level.
The peak column density ($N^{p}_\mathrm{tot}$) is measured within a 25\arcsec\ (10,000~au) circular aperture centered on the target position.
Note that the listed  $N^{p}_\mathrm{tot}$ may be overestimated due to nearby bright sources within the 25\arcs aperture. 
}
\tablehead{
\colhead{} & \colhead{} & \multicolumn{2}{c}{N$_2$H$^+$} & \multicolumn{2}{c}{CCS} & \multicolumn{2}{c}{CH$_3$OH} & \multicolumn{2}{c}{c-C$_3$HD} \\
\colhead{Name} & \colhead{Short Name} & \colhead{$\Delta v$} & \colhead{$N^{p}_\mathrm{tot}$} & \colhead{$\Delta v$} & \colhead{$N^{p}_\mathrm{tot}$} & \colhead{$\Delta v$} & \colhead{$N^{p}_\mathrm{tot}$}  \\
\colhead{} & \colhead{} & \colhead{(km s$^{-1}$)} & \colhead{(cm$^{-2}$)} & \colhead{(km s$^{-1}$)} & \colhead{(cm$^{-2}$)} & \colhead{(km s$^{-1}$)} & \colhead{(cm$^{-2}$)} & \colhead{(km s$^{-1}$)} & \colhead{(cm$^{-2}$)} 
}
\startdata
G203.21-11.20E1 & G203.21E1 & 1.97 & $5.6\times10^{12}$ & 1.95 & $1.3\times10^{12}$ & 2.27 & $1.9\times10^{13}$ & 2.05 & $1.2\times10^{12}$ \\
G205.46-14.56M3 & G205.46M3 & 2.56 & $1.1\times10^{13}$ & 1.17 & $8.9\times10^{11}$ & 3.41 & $1.8\times10^{13}$ & 1.36 & $3.3\times10^{11}$ \\
G206.21-16.17N  & G206.21N  & 1.77 & $6.4\times10^{12}$ & \nodata & \nodata & 2.64 & $6.9\times10^{12}$ & \nodata & \nodata \\
G206.21-16.17S  & G206.21S  & 1.18 & $7.8\times10^{12}$ & \nodata & \nodata & 0.94 & $3.8\times10^{12}$ & \nodata & \nodata \\
G208.68-19.20N2 & G208.68N2 & 2.75 & $2.8\times10^{13}$ & 0.97 & $8.4\times10^{11}$ & 2.65 & $2.4\times10^{13}$ & \nodata & \nodata \\
G209.29-19.65N1 & G209.29N1 & 2.94 & $8.4\times10^{12}$ & \nodata & \nodata & 2.46 & $8.6\times10^{12}$ & \nodata & \nodata \\
G209.29-19.65S1 & G209.29S1 & 3.93 & $1.3\times10^{13}$ & 0.98 & $5.1\times10^{11}$ & 4.54 & $1.6\times10^{13}$ & \nodata & \nodata \\
G209.29-19.65S2 & G209.29S2 & 3.92 & $9.8\times10^{12}$ & \nodata & \nodata & 3.59 & $2.4\times10^{13}$ & \nodata & \nodata \\
G209.55-19.68N2 & G209.55N2 & 1.57 & $3.9\times10^{12}$ & \nodata & \nodata & 1.32 & $1.1\times10^{13}$ & \nodata & \nodata \\
G209.77-19.40E3 & G209.77E3 & 1.77 & $2.8\times10^{12}$ & 0.78 & $5.1\times10^{11}$ & 1.71 & $1.3\times10^{13}$ & \nodata & \nodata \\
G209.94-19.52N  & G209.94N  & 1.96 & $1.3\times10^{13}$ & 1.56 & $7.1\times10^{11}$ & 2.08 & $3.3\times10^{13}$ & 2.38 & $9.6\times10^{11}$ \\
G210.37-19.53N  & G210.37N  & 1.37 & $4.0\times10^{12}$ & \nodata & \nodata & 1.89 & $8.3\times10^{12}$ & 1.70 & $4.9\times10^{11}$ \\
G210.82-19.47N2 & G210.82N2 & 1.38 & $4.6\times10^{12}$ & \nodata & \nodata & 1.70 & $1.9\times10^{13}$ & 1.36 & $6.2\times10^{11}$ \\
G211.16-19.33N4 & G211.16N4 & 1.57 & $4.7\times10^{12}$ & 1.56 & $2.2\times10^{12}$ & 2.65 & $2.0\times10^{13}$ & 1.71 & $1.3\times10^{12}$ \\
G211.16-19.33N5 & G211.16N5 & 2.16 & $9.3\times10^{12}$ & 1.17 & $1.2\times10^{12}$ & 1.89 & $1.6\times10^{13}$ & 2.05 & $5.6\times10^{11}$ \\
G212.10-19.15N1 & G212.10N1 & 1.97 & $6.1\times10^{12}$ & 2.33 & $6.3\times10^{12}$ & 2.64 & $2.2\times10^{13}$ & 2.04 & $2.8\times10^{12}$
\enddata
\end{deluxetable}

\clearpage\subsubsection{G203.21-11.20E1 (G203.21E1)}

This source was not detected in ALMA 12-m data of ALMASOP project \citet{2020Dutta_ALMASOP}. 
Another starless core, G203.21-11.20E2, is located at the east of this source with a separation of $\sim$2\arcs. 
The distribution of N$_2$H$^+$ resembles an isosceles acute triangle, with its acute angle pointing east and the two equal sides extending toward the north and south. 
A particularly strong molecular anti-correlation is observed in this source: CH$_3$OH is concentrated on the northern side, c-C$_3$HD resides on the southern side, and CCS appears as two distinct features located farther south, beyond the c-C$_3$HD emission region.

\clearpage\subsubsection{G205.46-14.56M3 (G205.46M3)}

The clump G205.46–14.56M3 was referred to as G205.46–14.56North1 in the JCMT survey reported by  \citet{2018Yi_PGCC_SCUBA2_II}.
Another protostellar core, G205.46–14.56M2 (also known as G205.46–14.56North2), which harbors HOPS-387 and HOPS-386, is located to the southeast, with a projected separation of approximately 30\arcs.
\citet{2021Sahu_ALMASOP_presstellar} reported that G205.46–14.56M3 is fragmented into two substructures (B1 and B2) on a scale of $\sim$1,000 au.
\citet{2025Lin_G205.46-14.56M3_oH2D+} detect o-H$_2$D$^+$ toward substructure B1 with an elevated N$_2$D$^+$/N$_2$H$^+$, tracing advanced deuteration. As demonstrated by \citet{2025Lin_G205.46-14.56M3_oH2D+}, A time-dependent, deuteration-focused chemo-dynamical model reproduces these observables if the core assembles on a free-fall timescale, suggestive of turbulent core fragmentation. 

The morphology of the molecular tracers in G205.46–14.56M3 is complex.
The N$_2$H$^+$ emission peaks slightly east of the continuum peak and extends toward the northwest, northeast, and south.
The CH$_3$OH, c-C$_3$H$_2$, and CCS emissions associated with the clump appear as two extended components located on the northeast and northwest sides of the core.
The two CH$_3$OH components exhibit comparable column densities; c-C$_3$H$_2$ is stronger in the northwest, while CCS is more prominent in the northeast.

\clearpage\subsubsection{G206.21-16.17N (G206.21N)}

Overall, this clump exhibits a north–south elongation, and the N$_2$H$^+$ emission is resolved into three features: northern, middle, and southern.
The southern feature corresponds to J054138.5–013621, a submillimeter source identified by \citet{2016Kirk_JCMT}, but it is not classified as a young stellar object.
The CH$_3$OH emission generally follows the distribution of N$_2$H$^+$, with the northern feature showing the strongest CH$_3$OH intensity. 
CCS and c-C$_3$HD are not detected in this clump. 

\clearpage\subsubsection{G206.21-16.17S (G206.21S)}

In this clump, the N$_2$H$^+$ emission exhibits a core-like morphology.
The CH$_3$OH emission peaks near the N$_2$H$^+$ peak and displays a tail extending toward the north.
Neither CCS nor c-C$_3$HD is detected in this clump.

\clearpage\subsubsection{G208.68-19.20N2 (G208.68N2)}

This clump, also known as OMC-3 MMS4, is located in the northern part of the OMC-3 region.
Several sources are present within this field, arranged from northwest to southeast: HOPS 92 (OMC3 MMS 2, G208.68-19.20N1), HOPS 91 (OMC3 MMS 3), HOPS 89, HOPS 88 (OMC3 MMS 5), and HOPS 87 (OMC3 MMS 6, G208.68-19.20N1) \citep{2016Furlan_HOPS, 2020Dutta_ALMASOP}.
\citet{2024Hirano_G208.68-19.20N2} reported an extremely dense and compact object embedded within this clump.

The N$_2$H$^+$ emission shows a strong peak associated with the starless core.
However, due to the complex and crowded environment, it is difficult to confidently associate the other molecular components with specific structures.
CH$_3$OH and CCS emissions appear to be located to the northeast of the clump, though their peak positions differ.
The strong CH$_3$OH emission at the southeast side is likely associated with the hot corino in HOP 87 (OMC3 MMS 6, G208.68-19.20N1) field \citet{2024Hsu_HOPS87}. 
No c-C$_3$HD emission is detected.

\clearpage\subsubsection{G209.29-19.65N1 (G209.29N1)}

This clump is elongated in the northeast–southwest direction.
The continuum emission appears to be fragmented into two features, located to the north and south. 
While the N$_2$H$^+$ emission does not show significant fragmentation, it exhibits a strong peak at the northern feature with a tentative tail extending southward.
CH$_3$OH is distributed on the eastern side of the southern feature.
Neither c-C$_3$HD nor CCS is detected.

\clearpage\subsubsection{G209.29-19.65S1 (G209.29S1)}

This source, along with another located to the southwest, G209.29–19.65S2 (G209.29S2), is among the targets of this study.
The N$_2$H$^+$ emission extends slightly to the north.
The associated CH$_3$OH appears to have a feature toward the south, while another localized CH$_3$OH peak is found at the northwestern boundary of the larger clump.
c-C$_3$HD is not detected near the core.
CCS exhibits a core-like feature at the northwest.

\clearpage\subsubsection{G209.29-19.65S2 (G209.29S2)}

This source, together with G209.29–19.65S1 (G209.29S1) located to the northeast, is one of the targets in this study.
Within the extended N$_2$H$^+$ structure, the emission peak appears slightly offset from the continuum peak.
CH$_3$OH, in contrast, is distributed toward the southwest of the N$_2$H$^+$ peak.
Neither c-C$_3$HD nor CCS is detected in the region neighboring this core.

\clearpage\subsubsection{G209.55-19.68N2 (G209.55N2)}

This source is elongated along the northwest–southeast direction.
To the northeast lies the protostellar core HOPS 12 \citep[G209.55–19.68N1][]{2016Furlan_HOPS,2020Dutta_ALMASOP}, where a hot corino was detected \citep{2022Hsu_ALMASOP}.
The color scale of the CH$_3$OH image is dominated by the emission associated with HOPS 12.
It is clear that the CH$_3$OH emission does not align with the N$_2$H$^+$ distribution.

\clearpage\subsubsection{G209.77-19.40E3 (G209.77E3)}

The N$_2$H$^+$ emission from this source is relatively weak, likely due to low gas density.
Consistently, the dust continuum is also only weakly detected.
The CH$_3$OH and CCS emissions are distributed toward the northwest and west sides, respectively.
c-C$_3$HD is not detected.

\clearpage\subsubsection{G209.94-19.52N (G209.94N)}

This clump is elongated in the northwest–southeast direction.
N$_2$H$^+$ traces the dust continuum beautifully.
CH$_3$OH is distributed on the northeast side of the clump.
c-C$_3$HD is also aligned with the N$_2$H$^+$ emission.
CCS is present on both the northwestern side and the southeastern corner of the clump, with a peak at the latter location.

\clearpage\subsubsection{G210.37-19.53N (G210.37N)}

The continuum and N$_2$H$^+$ emissions are slightly elongated in the northwest–southeast direction.
CH$_3$OH is distributed along the southeastern side of the clump.
c-C$_3$HD emission is aligned with the N$_2$H$^+$ structure, while CCS is not detected.

\clearpage\subsubsection{G210.82-19.47N2 (G210.82N2)}

The continuum and N$_2$H$^+$ emissions are slightly elongated in the northwest–southeast direction.
The CH$_3$OH emission is peaking at the east side of the clump. 
c-C$_3$HD emission is aligned with the N$_2$H$^+$ structure, while CCS is not detected.
The source at the northwest–southeast side is a protostellar core HOPS 156 \citep[G210.82-19.47N1][]{2016Furlan_HOPS,2020Dutta_ALMASOP}. 

\clearpage\subsubsection{G211.16-19.33N4 (G211.16N4)}

This clump has a filamentary structure oriented in the northwest–southeast direction.
The N$_2$H$^+$ peak is offset by approximately 10\arcs\ to the southwest of the dust continuum peak.
Both c-C$_3$HD and CCS also peak on the southwest side, though their peak positions differ slightly.
In contrast, CH$_3$OH peaks on the southeastern side of the clump.

\clearpage\subsubsection{G211.16-19.33N5 (G211.16N5)}

The strong source on the southwest side is the protostellar core HOPS 135 \citep{2016Furlan_HOPS}.
Despite the presence of HOPS 135, it is clear that N$_2$H$^+$ is tracing our target, G211.16–19.33N5.
CH$_3$OH appears to have two components on the east and north sides, peaking in the east.
Both c-C$_3$HD and CCS are weak, with c-C$_3$HD distributed to the west and CCS to the east.

\clearpage\subsubsection{G212.10-19.15N1 (G212.10N1)}

The continuum emission of this clump is fragmented into two substructures, located to the east and west.
The N$_2$H$^+$ emission exhibits a large diamond-shaped morphology encompassing both continuum components.
These two substructures lie on the northeastern side of the N$_2$H$^+$ distribution, while the N$_2$H$^+$ peak is located between them.
CH$_3$OH is distributed toward the northern and southern corners of the clump.
The southern CH$_3$OH component shows a clear anti-correlation with the N$_2$H$^+$ peak.
Both c-C$_3$HD and CCS are distributed on the southeastern side of the N$_2$H$^+$ structure.
The c-C$_3$HD emission appears to split into two features: one lies on the western side of the eastern continuum substructure, and the other on the southern side of the N$_2$H$^+$ emission.
CCS shows a single component, peaking on the eastern side of the eastern continuum substructure.


\bibliography{REFERENCE.bib}{}
\bibliographystyle{aasjournal}




\end{CJK*}
\end{document}